\newif\ifdraft\draftfalse{}
\newif\iflater\latertrue{}
\newif\ifafterdeadline\afterdeadlinetrue{}

\PassOptionsToPackage{dvipsnames}{xcolor}

\documentclass[acmsmall,screen]{acmart}

\setcopyright{cc}
\setcctype{by}
\acmDOI{10.1145/3808329}
\acmYear{2026}
\acmJournal{PACMPL}
\acmVolume{10}
\acmNumber{PLDI}
\acmArticle{251}
\acmMonth{6}
\acmSubmissionID{pldi26main-p622-p}
\received{2025-11-13}
\received[accepted]{2026-04-03}

\colorlet{cb-blue}{RoyalBlue}
\colorlet{cb-green}{ForestGreen}
\colorlet{cb-pink}{CarnationPink}
\colorlet{cb-red}{RedOrange}

\usepackage{subcaption}
\usepackage{wrapfig}
\usepackage{mathpartir}
\usepackage{listings}
\usepackage[scale=0.82]{FiraMono}
\usepackage{xspace}
\usepackage[inline]{enumitem}
\usepackage{multirow}
\usepackage[subtle]{savetrees}

\definecolor{dkred}{rgb}{0.7,0,0}
\definecolor{dkblue}{rgb}{0,0,0.7}
\definecolor{notered}{rgb}{0.85,0,0}
\definecolor{dkpurple}{HTML}{4e02eb}
\definecolor{dkgreen}{HTML}{006329}
\definecolor{teal}{HTML}{007982}
\definecolor{fuchsia}{HTML}{8C368C}

\definecolor{keywordcolor}{HTML}{5e81ac}      
\definecolor{commentcolor}{RGB}{0,128,0}      
\definecolor{stringcolor}{HTML}{a3be8c}     

\lstdefinelanguage{lean}{
  morekeywords={def, inductive, where, match, with, let, rec, theorem, lemma,
  axiom, if, then, else, fun, do, return, have, show, by, import, namespace,
  section, variable, mutual, partial, deriving, structure, class, instance,
  extends, open},
  sensitive=true,
  morecomment=[l]{--},
  morecomment=[s]{/-}{-/},
  morestring=[b]",
  literate={=>}{$\Rightarrow$}2 {::}{$::$}1 {|}{$\mid$}1,
}

\newcommand{\rawimpname}{Palamedes\xspace}
\newcommand{\impname}{{\sc \rawimpname}\xspace}

\newcommand{\mainbench}{\textsc{Main}\xspace}

\newcommand{\cobbbench}{\textsc{Cobb}\xspace}

\newcommand{\Gen}{{\mathsf{Gen}}}
\newcommand{\gen}[1]{{\color{cb-blue} \mathsf{#1}}}
\newcommand{\bind}{\ {\color{cb-blue} \mathbin{>{\mkern-10mu}>{\mkern-10mu}=}\ }}

\newcommand{\code}[1]{\lstinline[language=Lean,mathescape=true]!#1!}
\newcommand{\mypara}[1]{\smallskip\noindent\emph{#1.}\xspace}

\theoremstyle{definition}
\newtheorem{definition}{Definition}[section]

\theoremstyle{definition}
\newtheorem{informal-definition}{Informal Definition}

\theoremstyle{definition}

\newtheorem{lemma}{Lemma}[section]

\AtBeginDocument{%
  \providecommand\BibTeX{{%
\normalfont B\kern-0.5em{\scshape i\kern-0.25em b}\kern-0.8em\TeX}}}

\title{The Search for Constrained Random Generators}

\author{Harrison Goldstein}
\email{hgoldste@buffalo.edu}
\orcid{0000-0001-9631-1169}
\affiliation{%
  \institution{University at Buffalo, SUNY}
  \city{Buffalo}
  \state{New York}
  \country{USA}
}

\author{Hila Peleg}
\email{hilap@cs.technion.ac.il}
\orcid{0000-0002-0107-5659}
\affiliation{%
  \institution{Technion}
  \city{Haifa}
  \country{Israel}
}

\author{Cassia Torczon}
\email{ctorczon@seas.upenn.edu}
\orcid{0009-0003-6717-9586}
\affiliation{%
  \institution{University of Pennsylvania}
  \city{Philadelphia}
  \state{Pennsylvania}
  \country{USA}
}

\author{Daniel Sainati}
\email{sainati@seas.upenn.edu}
\orcid{0009-0003-4738-8524}
\affiliation{%
  \institution{University of Pennsylvania}
  \city{Philadelphia}
  \state{Pennsylvania}
  \country{USA}
}

\author{Leonidas Lampropoulos}
\email{leonidas@umd.edu}
\orcid{0000-0003-0269-9815}
\affiliation{%
  \institution{University of Maryland}
  \city{College Park}
  \state{Maryland}
  \country{USA}
}

\author{Benjamin C. Pierce}
\email{bcpierce@cis.upenn.edu}
\orcid{0000-0001-7839-1636}
\affiliation{%
  \institution{University of Pennsylvania}
  \city{Philadelphia}
  \state{Pennsylvania}
  \country{USA}
}


\begin{abstract}
  Among the biggest challenges in property-based testing
  (PBT) is the {\em constrained random generation problem}: given a
  predicate on program values, randomly sample from the set of all
  values, and only values, satisfying that predicate. Efficient solutions to this problem are critical, since the
  executable specifications used by PBT
  often have preconditions that input values must satisfy in order to be
  valid test cases, and satisfying values are often sparsely
  distributed.

  We propose a novel approach to this problem using
  deductive program synthesis. We present a set of synthesis rules,
  based on a denotational semantics of generators, that give rise to an automatic
  procedure for synthesizing
  correct generators. Our system handles recursive
  predicates by rewriting them as catamorphisms and then
  matching with appropriate anamorphisms; this is theoretically
  simpler than other approaches to synthesis for recursive functions, yet still
  extremely expressive.

  Our implementation, \impname, is an extensible library for the Lean
  theorem prover. The synthesis algorithm itself is built out of standard
  proof-search tactics, reducing implementation burden and allowing the algorithm
  to benefit from further advances in Lean proof automation.
\end{abstract}


\begin{document}
\maketitle

\section{Introduction}\label{sec:intro}

Property-based testing (PBT)~\cite{claessenQuickCheckLightweightTool2000}
aims to bridge the gap between traditional testing and heavier-weight formal
methods~\cite{wrennUsingRelationalProblems2021} by allowing developers to
automatically test software systems against formal specifications. PBT
has been used to find bugs in a wide range of real-world
software~\cite{bornholtUsingLightweightFormal2021,artsTestingAUTOSARSoftware2015,artsTestingTelecomsSoftware2006,HowWeBuilt2023},
but important challenges remain.

One key challenge is the {\em constrained random
generation problem}. %
Consider a classic example of the kind of property one might test with PBT:
\[ \forall\ x~t,\ \textsf{isBST}(t) \implies \textsf{isBST}(\textsf{insert}(x,~t)) \]
This says that, if a tree $t$ is a valid binary search tree (BST), then
inserting a new value $x$ into $t$ yields another valid BST. To test this
property, a PBT framework uses a program called a {\em generator}
to randomly sample thousands of values for $x$ and $t$ and check that

\begin{wrapfigure}[9]{r}{0.3\textwidth}
\vspace{-15pt}
\begin{lstlisting}
def isBST t lo hi :=
  match t with
  | leaf => true
  | node l x r =>
    lo <= x && x <= hi &&
    isBST l lo (x - 1) &&
    isBST r (x + 1) hi
\end{lstlisting}
  \caption{A recursive Lean predicate that checks if a tree is a BST.}\label{fig:bst-predicate}
\end{wrapfigure}

\noindent
the property holds for each pair of values.

\mypara{Constrained random generation}
Not every way of sampling $x$ and $t$ will be effective.  In
particular, the
property is vacuously true if $t$ is not a valid BST to start with,
i.e., if it fails to pass the predicate in \autoref{fig:bst-predicate}.
If the generator is not carefully designed, most trees will
be discarded as invalid. The vast majority of labeled binary trees
with more than one or two nodes are not
ordered in a way that makes them
valid BSTs, so only a few will actually be used to test
the insert function.
This motivates the \emph{constrained random generation problem}: to test a conditional property
effectively, we need a way to randomly sample from the set of all and only the values that satisfy its precondition.

A PBT expert might address this problem by writing a generator manually---e.g., the one
in \autoref{fig:bst-hand-written}---%
that produces BSTs by construction. This generator tracks the
range of values that may validly appear in a given sub-tree.
If the range is empty, \code{pure} creates a constant generator
that always returns \code{leaf}. Otherwise
\code{pick} is used to make a choice:
either generate a \code{leaf} (this allows
us to generate trees that aren't of maximum height) or generate
a \code{node} by selecting a value in the appropriate range and
recursively generating subtrees with truncated ranges.
Lean's \code{do}
notation sequences generators by 

\begin{wrapfigure}{l}{0.35\textwidth}
\begin{lstlisting}
def genBST lo hi :=
  if lo > hi then
    pure leaf
  else
    pick
      (pure leaf)
      (do
        let x <- choose lo hi
        let l <- genBST lo (x - 1)
        let r <- genBST (x + 1) hi
        pure (node l x r))
\end{lstlisting}
  \caption{A handwritten generator for binary search trees.}\label{fig:bst-hand-written}
  \Description{A handwritten generator for binary search trees.}
\end{wrapfigure}

\noindent
sampling from one generator, binding the sampled
value to a variable, and continuing as another generator.

Unfortunately, PBT novices can struggle
to come up with such generators from scratch. Indeed, a recent study of PBT
users~\cite{goldsteinPropertyBasedTestingPractice2024} found that even experts,
who can in principle write effective generators like \code{genBST},
still see writing generators as a distraction from other
testing tasks.
Consequently,
this problem has been studied extensively by programming languages researchers
over the
years.
Many of the approaches
revolve around some kind of search procedure:
start with a na\"ive
generator (e.g., one derived from type
information~\cite{xiaGenericrandom2021,mistaDerivingCompositionalRandom2021})
and prune its results ``online'' during generation to rule out
invalid values,
using techniques like
constraint solving~\cite{steinhofelInputInvariants2022,seidelTypeTargetedTesting2015}
reinforcement learning~\cite{reddyQuicklyGeneratingDiverse2020},
laziness~\cite{claessenGeneratingConstrainedRandom2015},
iterated Brzozowski derivatives~\cite{goldsteinParsingRandomness2022},
constrained logic programming~\cite{deweyAutomatedDataStructure2015},
and large language models~\cite{xiaFuzz4AllUniversalFuzzing2024}.

But searching for valid inputs can make generators
unusably slow.  Many PBT users run their tests often, in a tight
loop~\cite{goldsteinPropertyBasedTestingPractice2024}, and such
workflows demand very efficient generation. Moreover, we know that
generating valid inputs without expensive searching is
possible---the \code{genBST} generator
in \autoref{fig:bst-hand-written} does no searching at all!
If we could {\em automatically} build generators that don't search,
we could significantly improve testing-time performance while giving
users the convenience they want.

\mypara{The Constrained Generator Synthesis Problem}
Following this intuition, we propose a new {\em constrained generator synthesis
problem} that refines the familiar constrained random generation
problem by requiring solutions that build generators like \code{genBST} ``offline,''
rather than performing search at testing time.
A couple of existing tools, in particular
\textsc{QuickChick}~\cite{lampropoulosGeneratingGoodGenerators2017} and
\textsc{Cobb}~\cite{lafontaineWeveGotYou2025}
address this more specific version of the problem
(we discuss these
in detail in \autoref{sec:evaluation}), but in general this version of the
problem is under-explored.

We propose a novel approach to the constrained generator synthesis problem that
uses a {\em deductive synthesis} algorithm to search for generators that are
guaranteed---by construction---to produce valid values without searching at run
time.
%
The approach is based on a denotational semantics that characterizes the
values that a given generator can generate. To get a generator for a
property whose precondition is $\varphi$, we compute a generator $g$ satisfying
$\forall\ a,\ \ a \in \llbracket g \rrbracket \iff \varphi(a)$,
where $\llbracket g \rrbracket$ is the set of values the generator can generate.
In other words, $a$ can be generated by $g$ if and only if $\varphi(a)$ holds:
the generator is \emph{sound} and \emph{complete}.
Inspired by systems like
Synquid~\cite{polikarpovaProgramSynthesisPolymorphic2016} and SuSLik~\cite{polikarpovaStructuringSynthesisHeapmanipulating2019}, our synthesis
algorithm works by building a {\em proof} that there exists an
appropriate $g$ for a given
$\varphi$ by applying a series of proof rules, then extracts a proof
witness that embodies $g$ in executable form.
The resulting generators are correct by construction, since the
search constructs the proof as well.

Our basic proof rules align with common
{\em generator combinators}---the core functions used to build
generators by hand. To construct recursive generators for inductively
defined data structures,
we use {\em recursion schemes}; concretely,
we observe that many predicates that can be represented as a fold (or
catamorphism) have an associated generator represented as an unfold (or
anamorphism). The upshot is that we can deal with first- and second-order
primitive-recursive predicates
using pre-derived induction principles, rather than
relying on cyclic proofs~\cite{itzhakyCyclicProgramSynthesis2021}.

It is important to note that our synthesis procedure is guaranteed to produce
generators with the correct \emph{support} (i.e., the right set of values), but not necessarily optimal
\emph{distributions}. We have chosen to treat generators as nondeterministic
programs, focusing on the set of values that they can produce and ignoring the
probabilities used to produce them.
This is a significant simplification, but an important one. Rather than
try to do everything at once, we think synthesizing generators with provably
optimal support and then \emph{tuning} them is significantly more tractable and 
flexible than including tuning in the synthesis process.
Ongoing work on generator
tuning~\cite{tjoaTuningRandomGenerators2025, zhuangAriadne2026} has made
promising steps towards automating
this second stage of the process already.
See \autoref{sec:future}.

\mypara{\rawimpname}
We implement our synthesis algorithm as a library called \impname
for the Lean theorem prover;
it builds Lean generators from predicates expressed as Lean functions.
Working in a theorem prover brings us several benefits.
First, the proofs produced by our deductive synthesis procedure can be
checked by Lean's core checker; this ensures that, even if there
were mistakes in the synthesis procedure, any successfully synthesized
generator is guaranteed to produce exactly the desired set of values.
Next, working in Lean
means the implementation of the synthesis procedure can be simplified
by using a popular proof search tactic,
Aesop~\cite{limpergAesopWhiteBoxBestFirst2023}, to do much of the heavy lifting.
Finally, extending the algorithm with new primitives or search tactics is as
simple as proving a lemma or writing a macro.
Our synthesizer is already quite powerful---it can, for example,
synthesize generators for BSTs and well-typed simply-typed lambda calculus
(STLC) programs---and the set of predicates it can handle can continue to grow
modularly.

We evaluate \impname against two other methods for automatically obtaining
PBT generators.
\impname is considerably faster than the state-of-the-art approach to generator
synthesis, Cobb~\cite{lafontaineWeveGotYou2025}, completing benchmarks like
BST orders of magnitude faster.
We also compare \impname to
QuickChick~\cite{paraskevopoulouComputingCorrectlyInductive2022,lampropoulosGeneratingGoodGenerators2017},
which is able to automatically derive generators from inductive relations in
Rocq; \impname is slower than QuickChick but imposes a lighter specification
burden, and we discuss the trade-offs between the two approaches in detail.

\autoref{sec:overview} below outlines our synthesis procedure in the context of
the BST example we have been discussing.
The remainder of the paper
presents the following contributions:
\begin{itemize}
  \item We propose a system of deductive synthesis rules for correct
    generators (\autoref{sec:theory}), proven correct relative to a
    denotational semantics of
    generators.
  \item We extend the resulting synthesis algorithm to work for
    generators of recursive data types (\autoref{sec:folds}),
    using tools from the
    literature on recursion schemes to greatly simplify
    the synthesis process.
  \item We offer \impname{}, an implementation of our synthesis procedure
    embedded as a library in the Lean proof assistant
    (\autoref{sec:implementation}).
    This implementation strategy achieves performant synthesis that
    also maintains mechanized
    proofs of correctness, while borrowing key parts of the algorithm from Lean's
    existing infrastructure.
  \item We evaluate our approach on a suite of 32 benchmarks from the literature, demonstrating that
  our synthesizer is either faster or more flexible than state-of-the-art
  approaches (\autoref{sec:evaluation}).
\end{itemize}
We conclude by
discussing
limitations (\autoref{sec:limitations}) and related and future work
(\autoref{sec:related} and \autoref{sec:future}).

\section{Overview}\label{sec:overview}

In this section we illustrate the components of \impname{} with a running example:
synthesizing a generator for binary search trees.

\mypara{Synthesis for Simple Predicates}
Our deductive synthesizer produces a PBT generator by recursive proof search.
At every step, \impname attempts to match the current predicate (or
goal) with the conclusion of an {\em inference rule} like the ones in
\autoref{fig:infrules-purepick}. Upon matching a rule, the process then
continues with the premises of this rule.
Concretely, to generate values satisfying a predicate $P$, we start with the statement
{\small
\[
  \inferrule*[right=?]
  { }
  { {?} \ :\ \mathsf{Gen}_\alpha~P }
  \]}%
and successively refine the proof tree until we have a complete generator.
For example, given a predicate like $\lambda\ a \Rightarrow a = 1 \lor
a = 2$, the synthesizer would choose $\gen{pick}$ because the
disjunction in its conclusion matches the form of the predicate.
Its two component predicates, $\lambda\ a \Rightarrow a = 1$ and $\lambda\ a \Rightarrow a = 2$,
become new goals for the synthesizer, both of which are matched by the rule for
$\gen{pure}$ with no new further assumptions.
The final result is
generator
$\gen{pick}~(\gen{pure}~1)~(\gen{pure}~2)$.
We show a step-by-step derivation
capturing this process in \autoref{appendix:basic-generator-derivation}.

\begin{wrapfigure}[6]{r}{0.37\textwidth}
\vspace{-5px}
\centering
  {\small
    $
    \begin{array}{c}
      \inferrule*
      { }
      { \gen{pure}~a' \ :\ \mathsf{Gen}~(\lambda\ a \Rightarrow a = a') }
      \\\\
      \inferrule*
      { g_1 \ :\ \mathsf{Gen}~P \\ g_2 \ :\ \mathsf{Gen}~Q }
      { \gen{pick}~g_1~g_2 \ :\ \mathsf{Gen}~(\lambda\ a \Rightarrow P~a \lor Q~a) }
    \end{array}$
  }
  \caption{Synthesis rules for $\gen{pure}$ and $\gen{pick}$.}\label{fig:infrules-purepick}
\end{wrapfigure}

Matching inference rules to predicates is not always so
straightforward. For example, in the process of synthesizing a generator for
the \code{isBST} predicate in \autoref{fig:bst-predicate}
(in particular, in the sub-case generated by the \code{leaf}
branch of the match), the synthesizer arrives at the goal predicate $\lambda\ a
\Rightarrow \mathsf{true} \land a = \mathsf{leaf}$.
This is logically equivalent to $\lambda\ a \Rightarrow a =
\mathtt{leaf}$, which matches the conclusion of the $\gen{pure}$ rule, but is not
exactly the same.  In a stand-alone synthesizer, we would need to add an
inference rule for each such case, but in
\impname we can use
Lean's built-in theorems help us.  In this case, Lean can automatically simplify
the former predicate into the latter, allowing the synthesizer to apply
the $\gen{pure}$ rule.
The same concept means that \impname can apply the $\gen{pick}$ rule to choose between
the \code{leaf} and \code{node} cases of the \code{isBST} predicate. Lean
rewrites the \code{match} to a simple disjunction, which the synthesizer uses to
produce the \code{pick} that appears in \code{genBST}.

\begin{wrapfigure}[4]{r}{0.5\textwidth}
\centering
  {\small
    $\inferrule*
    {g \ :\ \mathsf{Gen}~P \\ f \ :\ (a' : \alpha') \to \mathsf{Gen}~(\lambda\ a \Rightarrow\ Q~a'~a) }
    {g \bind f \ :\ \mathsf{Gen}~(\lambda\ a \Rightarrow \exists\ (a' : \alpha'),\ P~a' \land Q~a'~a)}$}
  \caption{Synthesis rule for $\gen{bind}$ ($\bind$).}\label{fig:infrules-bind}
\end{wrapfigure}

A hallmark of PBT generators is the ability to
express dependencies between generated values. For example, \code{genBST}
generates \code{x} by sampling a \code{choose lo hi} to pick a value between
\code{lo} and \code{hi},
and it uses that $x$ later
in the generator. Our synthesizer
handles these dependencies with the rule in \autoref{fig:infrules-bind}.
The $\gen{bind}$ rule's conclusion matches a conjunction like the one in the
body of \code{isBST}, where \code{lo <= x && x <= hi} is conjoined with
\code{isBST l lo (x - 1) && isBST r (x + 1) hi}. When we apply the rule, we get
two new
obligations: one for a generator that can be sampled to give a value \code{x}
(in \code{isBST} that's \code{choose lo hi})
and another for a generator that uses \code{x} to continue generating.
(The binds that
produce values for \code{l} and \code{r} are a bit more complicated---we discuss
them in a moment when we explain how we handle recursion.)

\mypara{Recursive Predicates}
Recursion is challenging for deductive synthesizers because generating recursive programs
requires reasoning about termination. Previous
work~\cite{itzhakyCyclicProgramSynthesis2021,polikarpovaStructuringSynthesisHeapmanipulating2019}
generates terminating recursive calls by proving
that some measure decreases on each iteration, but generators often have much more
complicated termination arguments than standard functional programs.
Indeed, generators for some infinite sets (e.g., lists or trees) are often
written in a way that is not guaranteed to terminate at all: they
terminate with high probability but may have infinite execution traces.

\begin{figure}[h]
\begin{tabular}{p{0.55\textwidth}p{0.45\textwidth}}
\begin{lstlisting}
def genBSTUnfold lo hi :=
  Tree.unfold
    (fun (lo, hi) =>
      if lo > hi then
        pure leafStep
      else
        pick
          (pure leafStep)
          (do
            let x <- choose lo hi
            pure (nodeStep (lo, x - 1) x (x + 1, hi))))
    (lo, hi)
\end{lstlisting}
&
\begin{lstlisting}
def isBSTFold lo hi t :=
  Tree.fold
    (fun bl x br s =>
      match s with
      | (lo, hi) =>
        (lo <= x && x <= hi) &&
        bl (lo, x - 1) &&
        br (x + 1, hi))
    (fun _ => true)
    t
    (lo, hi)
\end{lstlisting}
\end{tabular}
\vspace{-0.5em}
\caption{Left: a version of \code{genBST} using \code{Tree.unfold}.
Right: a version of \code{isBST} using
\code{Tree.fold}.}\label{fig:bst-unfold}\label{fig:isbstfold}
\Description{A generator for binary search trees and the predicate that it
adheres to.}
\end{figure}

Luckily, it is possible to synthesize recursive generators without reasoning
directly about individual recursive calls. The trick is to use a {\em recursion
scheme}, and in particular an {\em unfold} (or anamorphism).  Recursion schemes
are popular in the functional programming
literature~\cite{yangFantasticMorphismsWhere2022,huttonTutorialUniversalityExpressiveness1999}
as a way of abstracting common patterns of recursion, and it turns out that they
work well for generators too. At a high level, an unfold describes a way to grow
a data structure from a seed value. It takes as input a single step of the
process, and then it ``ties the knot'' to turn that single step into a
fully-fledged generator. You can see an example of this in the
\code{genBSTUnfold} generator in \autoref{fig:bst-unfold}. The function passed
to \code{Tree.unfold} either produces a \code{leafStep}, which says that the
generator should produce a \code{leaf} and stop, or a \code{nodeStep}, which says
that the generator should produce a \code{node} by continuing the process again
with new seed values for the left and right sub-trees.  This process ultimately
does the same thing as \code{genBST} in \autoref{fig:bst-hand-written}.
We go into more detail on unfolds in 
\autoref{sec:folds}; for now, it suffices to understand that unfolds let us
synthesize generators for recursive generators without explicitly dealing with
recursive functions.

\begin{wrapfigure}[4]{r}{.45\textwidth}
  \centering
  {\small
   $\inferrule*
    { g \ :\ (b : \beta) \to \Gen~(\dots) }
    { \mathsf{Tree.unfold}~g~b \ :\ \mathsf{Gen}~(\lambda\ \mathit{t}\ \Rightarrow \mathsf{Tree.fold}~f~z~\mathit{t} = b) }$}
   \caption{Synthesis rule for $\mathsf{Tree.unfold}$.}\label{fig:infrules-unfold}
\end{wrapfigure}

How do we synthesize unfolds? By observing that unfolds
are dual to another kind of recursion scheme, the {\em fold} (or catamorphism).
Folds can express a wide
range of recursive algorithms;
in particular, folds can express
predicates like \code{isBST} (\code{isBSTFold} in \autoref{fig:isbstfold}).
The fold version of the predicate essentially does the unfold in reverse: rather
than take a seed value and expand it into a tree, it takes a tree and collapses
it into a value. This is the duality that makes our approach work.

Putting this all together,
we take a predicate like \code{isBST},
convert it to a form like \code{isBSTFold} that uses a fold, 
apply a rule like the one in \autoref{fig:infrules-unfold}, and then
continue with the deductive synthesis process outlined above
to obtain a generator that
looks like \code{genBSTUnfold}.
This whole pipeline is automated (see \autoref{sec:folds}), and means
that we can synthesize generators based on recursive predicates without actually
needing our synthesis procedure to understand recursion explicitly.

As described above, this approach works for first- and second-order primitive
recursive functions, i.e., those expressible with first- and second-order folds.
Luckily, many common examples of predicates---including complex ones like
filtering for well-typed programs---fit this description. In \autoref{sec:folds}
we describe our handling of recursive predicates in more detail.

\section{Deductive Synthesis for Generators}\label{sec:theory}

\impname proposes a novel approach to the constrained generator
synthesis problem. This section introduces its representation of generators and
some key definitions (\autoref{sec:theory:representation}), presents our
core deductive synthesis rules for constructing generators
(\autoref{sec:theory:basic}), introduces some handy base
generators
(\autoref{sec:theory:standardlib}), and describes an optimization procedure
that can be applied to generators after synthesis
(\autoref{sec:theory:optimization}).

\subsection{Generator Representation}\label{sec:theory:representation}

Generators for values of type \code{a} are often
represented as sampling functions of type
\code{Seed $\to$ a}, but we opt for a representation with a bit more
flexibility. Taking inspiration from work on {\em free
generators}~\cite{goldsteinParsingRandomness2022}, we represent a generator as
an inductive datatype (\autoref{fig:gen-support}) that can be
interpreted in multiple ways.

Each of these constructors has a standard
interpretation as a procedure for sampling values. This interpretation
is discussed in detail
in \autoref{sec:implementation}; the intuition is as
follows. The $\gen{pure}$ constructor
represents a constant generator that always produces the same value. The $\bind$
constructor represents sequencing of generators, sampling from one and
passing the sampled
value to a
function producing another. The $\gen{pick}$ constructor represents a
random choice
between generators.

\begin{wrapfigure}[10]{l}{.55\textwidth}
  {\small
  \vspace*{1ex}
\begin{lstlisting}
inductive Gen @$\alpha$@ where
  pure : @$\alpha$@  @$\to$@ Gen @$\alpha$@
  pick : Gen @$\alpha$@ @$\to$@ Gen @$\alpha$@ @$\to$@ Gen @$\alpha$@
  bind : Gen @$\alpha$@ @$\to$@ (@$\alpha$@ @$\to$@ Gen @$\beta$@) @$\to$@ Gen @$\beta$@
  indexed : (@$\mathbb{N}$@ @$\to$@ Gen (Option @$\alpha$@)) @$\to$@ Gen @$\alpha$@
  assume : (b : @$\mathbb{B}$@) @$\to$@ (b = true @$\to$@ Gen @$\alpha$@) @$\to$@ Gen @$\alpha$@
\end{lstlisting}
  }
  \caption{The datatype of generators. We use \code{>>=} or $\bind$ as an infix variant of \code{bind}.}\label{fig:gen-support}
  \vspace*{-1ex}
\end{wrapfigure}

The $\gen{indexed}$
constructor represents an infinite family of generators, indexed by natural
numbers; larger numbers give the generators more ``fuel'' to
produce more values.
(In lazy languages like Haskell this constructor is not
necessary. But since we will be working in Lean, which is strict, we need
it to be able to represent generators of infinite sets like natural
numbers and lists.)
The $\gen{assume}$ constructor represents a partial generator
parameterized on a boolean condition; if true, it simply calls its argument;
if false, it generates nothing. When interpreted as a sampling function,
it is a partial function that can fail and need to
be retried or backtracked around.

For proofs, we need an interpretation of generators that characterizes the
{\em set} of values that they can produce via sampling. This interpretation is
inspired by \citet{paraskevopoulouFoundationalPropertyBasedTesting2015}.
\begin{definition}\label{def:support}
  The {\em support} of a generator $g$ is the set of values that $g$ can
  produce, written $\llbracket g \rrbracket$.
  {\small
    \begin{align*}
      a \in \llbracket \gen{pure}~a' \rrbracket  &\iff a = a' \\
      a \in \llbracket x \bind f \rrbracket    &\iff \exists\ a',\ a' \in \llbracket x \rrbracket \land a \in \llbracket f~a' \rrbracket \\
      a \in \llbracket \gen{pick}~x~y \rrbracket &\iff a \in \llbracket x \rrbracket \lor a \in \llbracket y \rrbracket \\
      a \in \llbracket \gen{indexed}~f \rrbracket &\iff
      (\exists\ n,\ \mathsf{some}~a \in \llbracket f~n \rrbracket) \\
      &\qquad \land (\forall a~n,\ \mathsf{some}~a \in \llbracket f~n \rrbracket \to \mathsf{some}~a \in \llbracket f~(n + 1) \rrbracket) \\
      a \in \llbracket \gen{assume}~b~\gen{in}~x \rrbracket &\iff b = \mathsf{true} \land a \in \llbracket x \rrbracket
    \end{align*}
}
\end{definition}
\noindent
Note that the rule for $\gen{indexed}$ requires that the function passed to it be
{\em monotonic}; without this condition, the generator could produce fewer

\begin{wrapfigure}[12]{r}{0.5\textwidth}
  \vspace{-15px}
  {\small
\begin{lstlisting}
def arbNat : Gen @$\mathbb{N}$@ :=
  let rec go (fuel : @$\mathbb{N}$@) : Gen (Option @$\mathbb{N}$@) :=
    match fuel with
    | 0 => pure none
    | fuel' + 1 =>
      pick (pure (some 0))
            (go fuel' >>= fun on' =>
               match on' with
               | none => pure none
               | some n' => pure (some (1 + n')))
  indexed go
\end{lstlisting}
}%
\caption{A generator of natural numbers.}\label{fig:arbNat}
\end{wrapfigure}

\noindent values with more fuel, which would break completeness. Our synthesis procedure
enforces this condition by construction. This setup means that support is
not termination sensitive; a generator with support $P$ will produce all values
satisfying $P$ (terminating when it does so).

\mypara{Example: Generator for Natural Numbers}
\autoref{fig:arbNat} uses most of the above constructors to produce a generator
for natural numbers. It uses $\gen{indexed}$, since $\mathsf{go}$ is a
fuel-indexed recursive function.
If the fuel has run out,
$\mathsf{go}$ fails with $\mathsf{none}$. Otherwise, it
makes a random choice between  $0$ and  $1 + n'$,
where $n'$ is generated by recursively calling $\mathsf{go}$.
The support of this generator is precisely $\mathbb{N}$.

\mypara{Example: Backtracking Generator}
While our ultimate goal is to avoid generators that search during
generation, we still need such generators to be representable in our
language.
As we show in \autoref{sec:theory:optimization}, our synthesis procedure works by first producing generators
that need to backtrack and then attempting to optimize them.
Here is an example of a generator that backtracks due to $\gen{assume}$:
\begin{center}
  {\small
\code{def genBacktrack := pick (pure 1) (assume false (fun _ => pure 2))}
  }
\end{center}
The support of this generator is $\{1\}$---i.e., when it generates a
value, it always generates $1$---but it will sometimes choose the
right side of the $\gen{pick}$ and have to backtrack and try again.

\mypara{Generator Correctness}
We write $\mathsf{Gen}_\alpha~\varphi$ for the type of correct
generators with respect to a predicate $\varphi$ on a type $\alpha$,
eliding the latter if it is clear from context.
\begin{definition}[Correctness]
  A generator $g$ is {\em correct} with respect to a predicate $\varphi$ if it
  is both {\em sound}, i.e., $\forall\ a,\ a \in \llbracket g \rrbracket
  \implies \varphi(a),$ and {\em complete}, i.e.,
  $\forall\ a,\ \varphi(a) \implies a \in \llbracket g \rrbracket.$
\end{definition}
For example, the
$\mathsf{arbNat}$ generator above can be given the type
$\Gen_\mathbb{N}~(\lambda\ n \Rightarrow \top)$, while
$\mathsf{genBacktrack}$ has type $\Gen_\mathbb{N}~(\lambda\
n \Rightarrow n = 1)$.

\mypara{Assume-Freedom}
Our gold standard for generators is exemplified
by \code{genBST}---generators that behave like the ones written by
expert users to produce valid inputs by construction. The synthesis procedure we
describe over the next few sections meets this standard in most cases,
but there are situations in which it is not completely
successful.
In particular, it sometimes produces generators that use
the $\gen{assume}$ constructor in ways that require backtracking.
To distinguish these correct-but-suboptimal generators from the more
performant ones we aim to synthesize as often as possible, we define:

\begin{definition}[Assume-Freedom]\label{def:assume-freedom}
  A generator $g$ is {\em assume-free} iff it does not mention the
  $\gen{assume}$ constructor (including in sub-generators that it calls).
\end{definition}

\subsection{Core Synthesis Algorithm}\label{sec:theory:basic}
We now outline a (semi-)algorithm to solve the correct generator synthesis
problem. It uses {\em deductive program synthesis}, constructing a
generator by working backwards from the structure of a given
predicate, creating a
proof that witnesses the generator's correctness and building the generator
itself {\em en passant}. This approach to synthesis can also be found
in systems like
SuSLik~\cite{polikarpovaStructuringSynthesisHeapmanipulating2019} and
Synquid~\cite{polikarpovaProgramSynthesisPolymorphic2016}.

\begin{figure}[ht]
{\small
\begin{gather*}
  \inferrule*[right=S-Pure]
  { \Gamma \ \vdash\ a' \ :\  \alpha }
  { \Gamma \ \vdash\ \gen{pure}~a' \ :\  \mathsf{Gen}_\alpha~(\lambda\ a \Rightarrow a = a') }
  \qquad
  \inferrule*[right=S-Pick]
  { \Gamma \ \vdash\ g_1 \ :\  \mathsf{Gen}_\alpha~P \\ \Gamma \ \vdash\ g_2 \ :\ \mathsf{Gen}_\alpha~Q }
  { \Gamma \ \vdash\ \gen{pick}~g_1~g_2 \ :\  \mathsf{Gen}_\alpha~(\lambda\ a \Rightarrow P~a \lor Q~a) }
  \\
  \inferrule*[right=S-Bind]
  {\Gamma \ \vdash\ g \ :\  \mathsf{Gen}_{\alpha'}~P \\ \Gamma \ \vdash\ f\ :\ (a' : \alpha') \to \mathsf{Gen}_\alpha~(Q~a') }
  {\Gamma \ \vdash\ g \bind f \ :\  \mathsf{Gen}_\alpha~(\lambda\ a \Rightarrow \exists\ (a' : \alpha'),\ P~a' \land Q~a'~a)}
  \qquad
  \inferrule*[right=S-Convert]
  { \Gamma \ \vdash\ P = Q \\ \Gamma \ \vdash\ g \ :\  \mathsf{Gen}~Q }
  { \Gamma \ \vdash\ g \ :\  \mathsf{Gen}~P }
  \\
  \inferrule*[right=S-Assumption]
  { (x : \mathsf{Gen}_\alpha~P) \in \Gamma }
  { \Gamma \ \vdash\ x \ :\  \mathsf{Gen}_\alpha~P }
  \qquad
  \inferrule*[right=S-Intro]
  { b{:}\beta, \Gamma \ \vdash\ g\ :\ \mathsf{Gen}_\alpha~P }
  { \Gamma \ \vdash\ (\lambda\ b \Rightarrow g) \ :\  (b : \beta) \to
  \mathsf{Gen}_\alpha~P } \\
  \inferrule*[right=S-Uncurry]
  { \Gamma \ \vdash\ f \ :\  (b : \beta) \to (c : \gamma) \to \mathsf{Gen}_\alpha~(P~b~c) }
  { \Gamma \ \vdash\ (\lambda\ (b, c) \Rightarrow f~b~c) \ :\  (p : \beta \times \gamma) \to \mathsf{Gen}_\alpha~(P~(\mathsf{fst}~p)~(\mathsf{snd}~p)) }
  \\
  \inferrule*[right=S-SplitBool]
  { \Gamma \ \vdash\ g_t \ :\  \mathsf{Gen}_\alpha~(P~\mathsf{true}) \\
  \Gamma \ \vdash\ g_f \ :\  \mathsf{Gen}_\alpha~(P~\mathsf{false}) }
  { b{:}\mathbb{B}, \Gamma \ \vdash\ \mathsf{match}~b~\mathsf{with} \mid \mathsf{true} \Rightarrow g_t \mid \mathsf{false} \Rightarrow g_f\ :\ \mathsf{Gen}_\alpha~(P~b) }
  \\
  \inferrule*[right=S-SplitNat]
  { \Gamma \ \vdash\ g_z \ :\  \mathsf{Gen}_\alpha~(P~0) \\
  n'{:}\mathbb{N}, \Gamma \ \vdash\ g_s~n'\ :\ \mathsf{Gen}_\alpha~(P~(n' + 1)) }
  { n{:}\mathbb{N}, \Gamma \ \vdash\ \mathsf{match}~n~\mathsf{with} \mid 0 \Rightarrow g_z \mid n' + 1 \Rightarrow g_s~n'\ :\ \mathsf{Gen}_\alpha~(P~n) }
\end{gather*}}
\caption{\impname core synthesis rules.}\label{fig:core-rules}
\Description{Core synthesis rules for Palamedes.}
\end{figure}

\mypara{Pure and Pick}
We already saw the first two synthesis rules, \textsc{S-Pure} and
\textsc{S-Pick}, in \autoref{fig:infrules-purepick}.  The former says that we can
synthesize a value that is equal to a constant using $\gen{pure}$, and the
latter says that we can synthesize for a disjunction using $\gen{pick}$.

\mypara{Convert}
We also saw in \autoref{sec:overview} that synthesis rules may not apply
directly to the goal as stated. For example, the predicate
$\lambda\ a \Rightarrow \mathsf{true} \land a = \mathsf{leaf}$
{\em almost} matches the \textsc{S-Pick} rule, but
not quite.
In these situations, we rely on Lean's simplifier to convert the goal into a
form that matches one of our synthesis rules. Formally, this process is done via
the \textsc{S-Convert} rule in \autoref{fig:core-rules}.

\mypara{Assumptions and Functions}
If the generator we need is already available in the typing context $\Gamma$,
we can just use it via the \textsc{S-Assumption} rule.
When the synthesis goal is a function returning a generator, we can
apply an introduction rule.
The \textsc{S-Intro} says that if we can produce an appropriate generator given
$b{:}\beta$ in the context, then we can produce a (dependent) function from
$\beta$ to that generator.

We also need a special case for dealing with functions that take tuples as
arguments, which appear frequently as a result of our rules for recursive
functions (see \autoref{sec:folds}).  The \textsc{S-Uncurry} rule says that we
are free to synthesize a curried function and then uncurry it, if the goal is to
produce a function with a tuple argument.

\mypara{Bind}
We can use $\bind$ to chain generators together, as demonstrated in \autoref{sec:overview}:
{\small
\[
  \inferrule*
  {
    \inferrule*
    {\cdots}
    {\Gamma \ \vdash\ \cdots \ :\  \mathsf{Gen}~(\lambda\ j \Rightarrow j = 1 \lor j = 2)}
    \\
    \inferrule*
    {
      \inferrule*{ }{j{:}\mathbb{N},\Gamma \ \vdash\ \gen{pure}~(j + 3)\ :\ \mathsf{Gen}~(\lambda\ i \Rightarrow i = j + 3)}
    }
    {
      \Gamma \ \vdash\ \lambda\ j \Rightarrow \gen{pure}~(j + 3) \ :\  j{:}\mathbb{N} \to \mathsf{Gen}~(\lambda\ i \Rightarrow i = j + 3)
    }
  }
  {\Gamma \ \vdash\ \gen{pick}~(\gen{pure}~1)~(\gen{pure}~2) \bind (\lambda\ j \Rightarrow \gen{pure}~(j + 3)) \ :\
  \mathsf{Gen}~(\lambda\ i \Rightarrow \exists\ j,\ (j = 1 \lor j = 2) \land i = j + 3)}
\]}%
The goal here is to generate an $i$ that is equal to $j + 3$, where $j$ is
constrained to be either $1$ or $2$. To synthesize an appropriate generator, we
use {\sc S-Bind}, which gives two sub-goals. On the left, we need to
synthesize a generator for $j$, which we complete as above. On the
right, we apply {\sc S-Intro} and {\sc S-Pure} to produce the
continuation of the bind. The final generator generates $1$ or $2$ and then adds
$3$.

With the help of the {\sc Convert} rule, {\sc S-Bind} can apply in a
wide range of situations.  For example, the predicate
\code{isSome x} does not contain explicit existential
quantification, but it is equivalent to $\exists\ a,\ \top \land x = \mathsf{some}~a$,
so the synthesizer can {\sc Convert} the predicate and then apply {\sc S-Bind}.
We discuss the logical manipulations that are applied to predicates in
\autoref{sec:implementation:algorithm}.

\mypara{Case Splitting}
If a generator needs to do different things based on the value of a variable in
the context, it can use rules like \textsc{S-SplitBool} or \textsc{S-SplitNat}
to introduce pattern matches. Rules for inductive types other than booleans and
numbers can be derived from their
definitions (see \autoref{sec:implementation}).

\subsection{Standard Library Generators}\label{sec:theory:standardlib}
The rules from the previous section are the core of our synthesis algorithm, and
they can be used by themselves to build complex generators, but the real power of
our approach comes from its extensibility. Rather than ask the synthesis
process to synthesize generators for arbitrary predicates ``all the
way down,'' we can provide it with a library of
building blocks that it can assemble to build more complex generators.  For the
examples in the rest of this paper, we will need just a few such
building blocks; all are standard in PBT libraries.

\begin{figure}[ht]
  {\small
  \begin{gather*}
  \inferrule*[right=S-Choose]
  { \Gamma \ \vdash\ \mathit{lo} \leq \mathit{hi} }
  { \Gamma \ \vdash\ \mathsf{choose}~\mathit{lo}~\mathit{hi} \ :\  \mathsf{Gen}_\mathbb{N}~(\lambda a \Rightarrow \mathit{lo} \leq a \leq \mathit{hi}) }
  \\
  \inferrule*[right=S-ChoosePartial]
  { }
  { \Gamma \ \vdash\ \gen{assume}~\mathit{lo} \leq \mathit{hi}~\gen{in}~\mathsf{choose}~\mathit{lo}~\mathit{hi} \ :\  \mathsf{Gen}_\mathbb{N}~(\lambda v \Rightarrow \mathit{lo} \leq v \leq \mathit{hi}) }
  \\
  \inferrule*[right=S-Elem]
  { \Gamma \ \vdash\ \mathit{xs} \neq [] }
  { \Gamma \ \vdash\ \mathsf{elements}~\mathit{xs} \ :\  \mathsf{Gen}_\alpha~(\lambda a
  \Rightarrow a \in \mathit{xs}) }
  \\
  \inferrule*[right=S-ElemPartial]
  {  }
  { \Gamma \ \vdash\ \gen{assume}~\mathit{xs} \neq []~\gen{in}~\mathsf{elements}~\mathit{xs} \ :\  \mathsf{Gen}_\alpha~(\lambda a \Rightarrow a \in \mathit{xs}) }
  \end{gather*}}
  \caption{\impname standard library synthesis rules.}\label{fig:stdlib-rules}
  \Description{Library synthesis rules for Palamedes.}
\end{figure}

\mypara{Choose}
The $\mathsf{choose}$ generator picks a natural number in a defined range. We define it
by recursion:
\begin{center}
{\small
\code{def choose (lo hi :\ $\mathbb{N}$) := if lo = hi then pure lo else pick (pure lo) (choose (lo + 1) hi)}
}%
\end{center}
\begin{lemma}[Choose Support]
  If $\mathit{lo} \leq \mathit{hi}$, then
  $a \in \llbracket \mathsf{choose}~\mathit{lo}~\mathit{hi} \rrbracket \iff \mathit{lo} \leq a \leq \mathit{hi}$.
\end{lemma}
\noindent The corresponding synthesis rule, \textsc{S-Choose}, appears in \autoref{fig:stdlib-rules}.

In a synthesis context, it may not always be easy to show that $\mathit{lo} \leq
\mathit{hi}$---it may not even be true. This motivates a second
way to synthesize $\mathsf{choose}$ that checks its precondition dynamically, the
\textsc{S-ChoosePartial} rule.
Both rules are valid, but the second introduces the potential for the generator to fail:
if it is called with $\mathit{lo} > \mathit{hi}$, it will not be
able to produce a value. This generator is
sound in the sense that, if it produces a value, then that value satisfies the
given condition---but it is not ideal for testing. Luckily, we can usually optimize
the $\gen{assume}$ away later (see below).

\mypara{Elements}
The $\mathsf{elements}$ generator picks a random value from a
list:
\begin{center}
{\small
\code{def elements xs := match xs with | [x] => pure x | x :: xs' => pick (pure x) (elements xs')}
}%
\end{center}
\begin{lemma}[Elements Support and Synthesis]
  If $\mathit{xs} \neq []$, then
  $a \in \llbracket \mathsf{elements}~\mathit{xs} \rrbracket \iff a \in \mathit{xs}$.
\end{lemma}

\mypara{Greater Than and Less Than}
Finally, there are $\mathsf{greaterThan}$ and $\mathsf{lessThan}$
generators that take a single
natural number and can generate any number respectively greater or less than
that number.
They are similar to
$\mathsf{choose}$, so we defer them to
\autoref{appendix:greater-than-less-than}.

\subsection{Optimizing Generators to Avoid Assumes}\label{sec:theory:optimization}

In most cases where the synthesizer inserts assumptions, they can be optimized
away. For example, consider the following synthesized generator:
{\small
\[
  \inferrule*
  {
    \cdots
    \\
    \inferrule*{ }{\mathit{lo}{:}\mathbb{N}, \mathit{hi}{:}\mathbb{N} \ \vdash\ \gen{assume}~lo \leq hi~\gen{in}~\mathsf{choose}~\mathit{lo}~\mathit{hi}\ :\ \mathsf{Gen}~(\lambda\ a \Rightarrow \mathit{lo} \leq a \leq \mathit{hi})}
  }
  {\mathit{lo}{:}\mathbb{N}, \mathit{hi}{:}\mathbb{N} \ \vdash\ \gen{pick}~(\gen{pure}~0)~(\gen{assume}~\mathit{lo} \leq \mathit{hi}~\gen{in}~\mathsf{choose}~\mathit{lo}~\mathit{hi})\ :\ \mathsf{Gen} (\lambda\ a \Rightarrow a = 1 \lor \mathit{lo} \leq a \leq \mathit{hi})}
\]}%
Logically, this $\gen{assume}$ is not necessary. As written, the generator
makes a choice and then fails if it happened to choose the right branch and $\mathit{lo}
> \mathit{hi}$.  But it could just as well check $\mathit{lo} \leq \mathit{hi}$ first and only
choose the right branch if the check succeeds. Generalizing this
observation, we design a set of {\em optimization rules} that
rewrite generators to
avoid failures; these appear in \autoref{fig:optrules}. The rule we need for the
above case is rule (6), which
rewrites a $\gen{pick}$ containing an $\gen{assume}$ to an if statement
that checks the assumption before the choice.

\begin{figure}[t]
  {\small
\begin{align}
  \gen{pure}~v \bind{} f &\quad\rightsquigarrow\quad f~v \\
  (x \bind{} g) \bind{} f &\quad\rightsquigarrow\quad x \bind{} (\lambda a \Rightarrow g~a \bind{} f) & \\
  (\gen{assume}~b~\gen{in}~x) \bind{} f &\quad\rightsquigarrow\quad \gen{assume}~b~\gen{in}~(x \bind{} f) & \\
  x \bind{} (\lambda\ a \Rightarrow \gen{assume}~b~\gen{in}~(f a)) &\quad\rightsquigarrow\quad \gen{assume}~b~\gen{in}~(x \bind{} f) & \text{if}\ a \notin \mathsf{fv}(b) \\
  \gen{pick}~(\gen{assume}~b~\gen{in}~x)~y &\quad\rightsquigarrow\quad \mathsf{if}~b~\mathsf{then}~\gen{pick}~x~y~\mathsf{else}~y & \\
  \gen{pick}~x~(\gen{assume}~b~\gen{in}~y) &\quad\rightsquigarrow\quad \mathsf{if}~b~\mathsf{then}~\gen{pick}~x~y~\mathsf{else}~x &
\end{align}
}
\caption{Optimization rules for generators.
Rules (1) and (2) are standard monad equivalences, (3) and (4) describe
how $\gen{assume}$s interact with $\gen{bind}$s, and (5) and (6)
actually lift $\gen{assume}$s out of choices.
}\label{fig:optrules}
\Description{Optimization rules for generators.}
\end{figure}

\begin{lemma}[Optimizations Correct]
  Rules (1)--(6) do not change the support of the generator.
\end{lemma}

These rules are not complete: they may still leave $\gen{assume}$s in the
generator.  For example, a $\gen{pick}$ with $\gen{assume}$s on both
sides will still fail if both conditions are false.
In some of these cases, a stronger optimizer may be able
remove more assumes, but most $\gen{assume}$s left after
optimization indicate fundamental backtracking in the algorithm that the
synthesis process found.
Happily, for the vast majority of the
examples in \autoref{sec:evaluation}, they can derive a generator that
is assume-free.

This optimization procedure is critical for some of our most interesting
examples, including the BST example that we introduced in
\autoref{sec:overview}. When synthesizing a generator for BSTs, \impname uses
\textsc{S-ChoosePartial} to pick a value for a node that is within the correct
range, and then later the optimizer lifts the bounds check out of the $\gen{assume}$
and into a less expensive check. This is why the generator that we synthesize
for BST (roughly \code{genBSTUnfold} in \autoref{fig:bst-unfold}) checks
\code{lo > hi}.

\section{Synthesizing Generators for Recursive Predicates}\label{sec:folds}

We next describe how \impname handles recursive predicates over
data structures like lists and trees. Our approach is based on {\em recursion
schemes}, so we start with some background on
those~(\autoref{sec:folds:background}). Then we outline our synthesis procedure
in stages, describing the basic approach~(\autoref{sec:folds:simple-folds}),
adapting that approach to a wider range of
predicates~(\autoref{sec:folds:accu}), adding support for multiple conjoined
recursive predicates~(\autoref{sec:folds:tupling}), and finally putting
everything together~(\autoref{sec:folds:ergonomics}).

\begin{figure}[h]
  \includegraphics[width=0.9\textwidth]{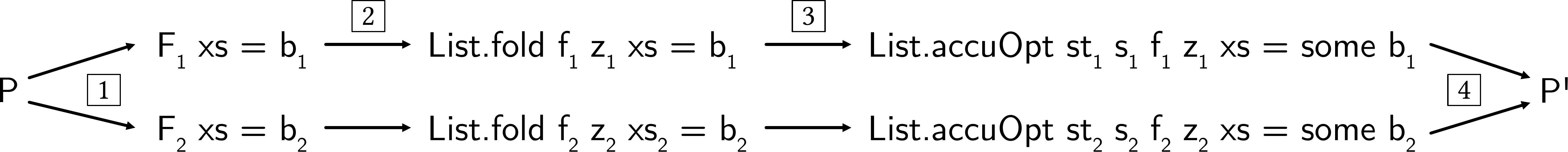}
  \caption{How a list predicate is normalized for synthesis.}\label{fig:fold-workflow}
  \Description{A cascading flow chart, starting from a function F and ending in
  a generator written with List.unfold.}
\end{figure}

At a high level, the approach in this section takes a recursive predicate,
normalizes it, and then applies a synthesis rule like the
ones from the previous section. The pipeline that we implement is shown in
\autoref{fig:fold-workflow}. The key takeaway is that, rather than give the synthesizer direct access to
$\gen{indexed}$ and recursion, we synthesize recursive generators through
higher-level rules. This approach does have limitations (e.g.,
it cannot create
generators like $\mathsf{elements}$ that iterate over one
structure and produce another), but it is highly effective for many predicates
over data structures.

\subsection{Background: Recursion Schemes}\label{sec:folds:background}

\begin{wrapfigure}[16]{r}{0.47\textwidth}
  \vspace{-10px}
{\small
\begin{lstlisting}
def List.fold
    (f : @$\alpha$@ @$\to$@ @$\beta$@ @$\to$@ @$\beta$@) (z : @$\beta$@)
    (xs : List @$\alpha$@) : @$\beta$@ :=
  match xs with
  | [] => z
  | x :: xs' => f x (fold f z xs')
\end{lstlisting}
\begin{lstlisting}
def List.accu
    (st : @$\alpha$@ @$\to$@ @$\sigma$@ @$\to$@ @$\sigma$@) (s : @$\sigma$@)
    (f : @$\alpha$@ @$\to$@ @$\beta$@ @$\to$@ @$\sigma$@ @$\to$@ @$\beta$@) (z : @$\sigma$@ @$\to$@ @$\beta$@)
    (xs : List @$\alpha$@) : @$\beta$@ :=
  match xs with
  | [] => z s
  | x :: xs' => f x (accu st (st x s) f z xs')
\end{lstlisting}
}
\caption{Definitions of folds and accumulators.}\label{fig:fold-and-acc}
\end{wrapfigure}

Recursive functions are a common challenge for program analysis and synthesis
tools, even in strongly normalizing languages where they are guaranteed to
terminate. While there are some existing techniques for synthesizing recursive
programs directly from recursive
specifications~\cite{polikarpovaStructuringSynthesisHeapmanipulating2019,polikarpovaProgramSynthesisPolymorphic2016},
we adopt a different approach that is simpler and easier to embed in Lean.

The functional programming community has produced a rich literature on {\em
recursion schemes}. Rather than express recursive functions directly via
unstructured general recursion, recursion schemes abstract recursion into
structured forms
that are easier to reason about.

\mypara{Folds}
The simplest recursion scheme is a {\em fold} or
{\em catamorphism}. \autoref{fig:fold-and-acc} shows an implementation of $\mathsf{fold}$ for the $\mathsf{List}$
datatype.
This function takes as arguments a ``base-case'' $z$ and a ``step function''
$f$. We call the type $\beta$ the ``collector'' for the
fold.\footnote{Others call this value the ``accumulator,'' but we use
  ``accumula\underline{tion}'' to
  refer to a type of fold~\cite{pardoGenericAccumulations2003} and want to
avoid confusion.} When the list is empty, we return $z$. When the list
is a cons cell, we
recursively call $\mathsf{List.fold}~f~z$ on its tail and use
$f$ to combine the resulting value with the value at the head.

Note that information here flows backward, from the tail of the list to the
head.\footnote{I.e., this is a ``right fold'' over the list
  ($\mathsf{List.foldr}$). In this paper we drop the ``r'' for
  consistency across  data structures: left folds are natural for
  lists, but they do not have an
analog for algebraic data types with branching recursion like trees.} We first
compute something about the tail, without considering the value at
the head, and only at the end do we actually put the two together. This point will be useful to remember.

\begin{wrapfigure}[10]{l}{.33\textwidth}
  \vspace{-8px}
\begin{minipage}{.33\textwidth}
\begin{lstlisting}
def isSorted xs :=
  List.accu
    (fun x _ => x)
    0
    (fun x b lo => lo <= x && b)
    (fun _ => true)
    xs
\end{lstlisting}
\end{minipage}
\caption{Checking if a list is sorted with \textsf{List.accu}.}\label{fig:accu-sorted}
\end{wrapfigure}

\mypara{More Advanced Recursion Schemes}
While $\mathsf{List.fold}$ can represent all primitive-recursive functions
over lists~\cite{huttonTutorialUniversalityExpressiveness1999}, we will see that
it is not always ergonomic to use it. For this reason, researchers
have identified specialized
recursion schemes, each capturing some common pattern of recursion~\cite{yangFantasticMorphismsWhere2022}. Of note here is the
{\em accumulation} pattern~\cite{pardoGenericAccumulations2003}, which
we present with some minor simplifications in \autoref{fig:fold-and-acc}.
Accumulations are similar to folds, but they pass information in both
directions. The collector value of type $\beta$ is still passed backwards through the
list, but we add a new type parameter $\sigma$ representing an ``accumulation
state'' that flows forward. The accumulation takes an initial state $s$ as input,
along with a ``state update function'' $\mathit{st}$ that says how the state
changes.

\begin{figure}[t]
\begin{minipage}{0.44\textwidth}
{\small
\begin{lstlisting}
    def List.foldOpt
          (f : @$\alpha$@ @$\to$@ @$\beta$@ @$\to$@ Option @$\beta$@)
          (z : Option @$\beta$@)
          (xs : List @$\alpha$@) : Option @$\beta$@ :=
      match xs with
      | [] => z
      | x :: xs' =>
        match List.foldOpt f z xs' with
        | none => none
        | some b' => f x b'
    \end{lstlisting}
}
\end{minipage}
\begin{minipage}{0.55\textwidth}
{\small
\begin{lstlisting}
  def List.accuOpt
        (st : @$\alpha$@ @$\to$@ @$\sigma$@ @$\to$@ @$\sigma$@) (s : @$\sigma$@)
        (f : @$\alpha$@ @$\to$@ @$\beta$@ @$\to$@ @$\sigma$@ @$\to$@ Option @$\beta$@)
        (z : @$\sigma$@ @$\to$@ Option @$\beta$@)
        (xs : List @$\alpha$@) : Option @$\beta$@ :=
    match xs with
    | [] => z s
    | x :: xs' =>
      match List.accuOpt st (st x s) f z xs' with
      | none => none
      | some b' => f x b' s
\end{lstlisting}
}
\end{minipage}
\caption{Definitions of \emph{optional} folds and
accumulators.}\label{fig:fold-and-acc-opt}
\Description{Optional folds and accumulators.}
\end{figure}

We can use $\mathsf{List.accu}$ to naturally implement some functions that would be
awkward with $\mathsf{List.fold}$. For example, the function in \autoref{fig:accu-sorted}
checks if a list of natural numbers is sorted.
The accumulation state is the minimum value allowed in the remainder
of the list; it is initialized to $0$ and updated to the most recently seen
value at each step.  The step function then checks that $\mathit{lo} \leq x$
and conjoins that with the boolean computed from the tail of the list,
to
ensure that the tail of the list is also sorted. We could implement this same
function with $\mathsf{List.fold}$,
but this would require the collector to be a higher-order
function, which would be difficult for the synthesizer to deal with.

\mypara{Optional Folds}
We also need versions of $\mathsf{List.fold}$ and
$\mathsf{List.accu}$ that capture {\em optional computations}
(\autoref{fig:fold-and-acc-opt}).
We discuss use cases for these in the next section; for now,
notice that they behave the same as $\mathsf{List.fold}$ and
$\mathsf{List.accu}$, except that, if any step
evaluates to $\mathsf{none}$, then the whole thing does.

\begin{wrapfigure}[16]{r}{.5\textwidth}
\vspace{-8px}
\begin{minipage}{.5\textwidth}
{\small
\begin{lstlisting}
def List.unfold
    (g : @$\beta$@ @$\to$@ Gen (Option (@$\alpha$@ @$\times$@ @$\beta$@))) (b : @$\beta$@) :
    Gen (List @$\alpha$@) :=
  let rec go b fuel :=
    match fuel with
    | 0 => pure none
    | 1 + fuel' =>
      g b >>= fun step =>
      match step with
      | none => pure (some [])
      | some (x, b') =>
        go b' fuel' >>= fun mxs =>
        pure (Functor.map (x :: @$\cdot$@) mxs)
  indexed (go b)
\end{lstlisting}
}
\end{minipage}
\caption{Definition of unfold for lists.}\label{fig:unfold-defn}
\end{wrapfigure}

\mypara{Unfolds}
The final recursion scheme we examine in detail is the {\em unfold} or {\em
anamorphism}. Unfolds are the inverse of folds: whereas folds collapse data
structures into compact values, unfolds expand values into data structures.
The
unfold for lists is shown in \autoref{fig:unfold-defn}.

The internal function $\mathsf{go}$ takes a seed value $b$ and some fuel.
If the
fuel is gone, it returns $\mathsf{none}$. Otherwise, it samples $g~b$
to obtain a ``$\mathsf{step}$''---if the $\mathsf{step}$ is $\mathsf{none}$,
generation terminates with an empty list, and, if the $\mathsf{step}$ is
$\mathsf{some}~(x, b')$, generation continues with a node containing the
value $x$ and a new seed value $b'$.  Finally, the $\gen{indexed}$
constructor
unifies this family of generators into a single generator for lists.

A key benefit of $\mathsf{List.unfold}$ is that it is guaranteed to make
exactly one recursive call for each element of the list it
produces. This means generators implemented with
$\mathsf{List.unfold}$ (as opposed to arbitrary general recursion) are
guaranteed to be efficient as long as their step functions are efficient.

\begin{wrapfigure}[10]{r}{.5\textwidth}
  \vspace{-23px}
\begin{minipage}{.5\textwidth}
{\small
\begin{lstlisting}
def isLengthK (k : @$\mathbb{N}$@) (xs : List @$\mathbb{N}$@) :=
  List.fold (fun _ b => 1 + b) 0 xs = k
def genLengthK (k : @$\mathbb{N}$@) : Gen (List @$\mathbb{N}$@) :=
  List.unfold
    (fun n =>
      match k with
      | 0 => pure none
      | 1 + k' =>
        arbNat >>= fun x =>
        pure (some (x, k'))) n
\end{lstlisting}
}
\end{minipage}
\caption{A predicate and generator for lists of length $k$.}\label{fig:lengthk-example}
\end{wrapfigure}

\subsection{Generators for Inductive Data Types}\label{sec:folds:simple-folds}
We now show how the tools for structuring recursion that we
reviewed in the previous subsection allow our synthesis procedure to
handle predicates over inductive data structures.

As a first example, given a predicate that uses
$\mathsf{List.fold}$ to check that a
list has a given length, we can
use $\mathsf{List.unfold}$ to write a generator for values satisfying this
predicate (\autoref{fig:lengthk-example}).
At each unfolding step, it checks the seed value $n$. If $n = 0$,
it generates $\mathsf{none}$, indicating that the list should end
(since $n$ is the target length of the list). Otherwise, it
generates an arbitrary natural number $x$ and yields $\mathsf{some}~(x, n -
1)$ to indicate that the list should continue with a cons cell containing $x$,
plus a decremented target length.

How might we derive $\mathsf{genLengthK}$ from
$\mathsf{isLengthK}$? The key observation is that
$\mathsf{isLengthK}$ and $\mathsf{genLengthK}$ have an inverse
relationship---whenever $\mathsf{genLengthK}$ takes a step, it is guaranteed
that $\mathsf{isLengthK}$ can undo that step.

\begin{lemma}[Fold-Unfold-Inverse for Lists]\label{lemma:fold-unfold}
  If, for all values $b$, the following relationship holds between an unfold's
  step function $g$ and a fold's arguments $f$ and $z$,
  \[
    \mathsf{none} \in \llbracket g~b \rrbracket \iff b = z \qquad\qquad
    \forall\ x~b',\ \mathsf{some}~(x, b') \in \llbracket g~b \rrbracket \iff b = f~x~b'
  \]
  then
  {
  $
    \forall\ \mathit{xs},\ \mathit{xs} \in \llbracket \mathsf{List.unfold}~g~b \rrbracket \iff \mathsf{List.fold}~f~z~\mathit{xs} = b
  $}.
\end{lemma}
The fold and
unfold are inverses here because, for each step the unfold takes, the fold is
guaranteed to be able to ``fold that step back up'' and recover the seed. Another perspective comes
from the observation that a fold passes information backwards in a list from the
tail to the head; the unfold does the opposite, passing information forwards
in such a way that the fold would always compute the same thing going the
other way.

We can use \autoref{lemma:fold-unfold} to prove the following synthesis
rule correct:
{\small
\[
  \begin{gathered}
    \inferrule*[right=S-List-Unfold']
    { \Gamma \ \vdash\ g \ :\  (b : \beta) \to \Gen_{\mathsf{Option}~(\alpha \times \beta)}~(P~b) }
    { \Gamma \ \vdash\ \mathsf{List.unfold}~g~b \ :\  \mathsf{Gen}_\mathbb{\mathsf{List}~\alpha}~(\lambda\ \mathit{xs}\ \Rightarrow \mathsf{List.fold}~f~z~\mathit{xs} = b) } \\
    \text{where} \quad P~b = \lambda\ \mathit{step} \Rightarrow (\mathit{step} = \mathsf{none} \land z = b) \lor (\exists\ x\ b',\ \mathit{step} = \mathsf{some}~(x, b') \land f~x~b' = b) \\
  \end{gathered}
\]}%
Our system can use this rule
to synthesize $\mathsf{genLengthK}$ from the definition of
$\mathsf{isLengthK}$.

\subsection{Handling More Complex Folds}\label{sec:folds:accu}
The {\sc S-List-Unfold'} rule is key to understanding the basics of our approach,
but it is not quite powerful enough for our most interesting use-cases.
For example, consider a predicate that checks that all
elements of a list are equal to $2$:
{\small
\begin{center}
\code{def isAllTwo (xs : List\ $\mathbb{N}$) := List.fold (fun x b => x = 2 && b) true xs = true}
\end{center}
}%
Using {\sc S-List-Unfold'}, we could turn this directly into a generator that
looks like this:
{\small
\begin{center}
\code{List.unfold (fun b => if b then pick (pure none) (pure (some (2, true))) else ...)}
\end{center}
}%
\noindent The $\mathsf{true}$ branch here is exactly what we want; it chooses
between \code{none}, which stops generation, or \code{some (2, true)} which says
to put a value \code{2} at the head of the list and continue.  But
$\mathsf{false}$ branch is unnecessary; it will never be executed, since $b$
starts as $\mathsf{true}$ and remains $\mathsf{true}$ every step through the
unfold. We would prefer to avoid synthesizing the $\mathsf{false}$ branch at
all.

The $\mathsf{isAllTwo}$ predicate has a hidden invariant that the
{\sc S-List-Unfold'} cannot make use of: if the step function ever returns
$\mathsf{false}$, the whole fold returns $\mathsf{false}$. We can make this
invariant available to the synthesizer by reinterpreting $\mathsf{isAllTwo}$ as an
optional fold:
{\small
\begin{center}
\code{List.foldOpt (fun x () => if x = 2 then some () else none) (some ()) xs = some ()}
\end{center}
}%
Now $\beta$ is $\mathsf{Unit}$, and the step function simply checks if $x = 2$
and, if not, fails. The invariant we wanted falls out of the definition of
$\mathsf{List.foldOpt}$.

As we saw earlier, $\mathsf{List.foldOpt}$ allows the fold to pass information
backward from the tail of the list to the front, but it does not allow state
to flow forward. We can be one step more general and consider predicates
expressed with $\mathsf{List.accuOpt}$.
This leads to our most general synthesis rule for recursive
predicates: the {\sc S-List-Unfold} rule, which subsumes and replaces {\sc
S-List-Unfold'}:
{\small
\setlength{\jot}{0pt}
\[
  \begin{gathered}
    \inferrule*[right=S-List-Unfold]
    { \Gamma \ \vdash\ g \ :\  (b : \beta) \to (s : \sigma) \to \Gen_{\mathsf{Option}~(\alpha \times (\sigma \times \beta))}~(P~b~s) }
    { \Gamma \ \vdash\ \mathsf{List.unfold}~g'~(b, s) \ :\  \mathsf{Gen}_\mathbb{\mathsf{List}~\alpha}~(\lambda\ \mathit{xs}\ \Rightarrow \mathsf{List.accuOpt}~\mathit{st}~s~f~z~\mathit{xs} = \mathsf{some}~b) } \\
    \begin{aligned}\\[-9px]
      \text{where} \quad &P~b~s = \\
      &\quad\lambda\ \mathit{step} \Rightarrow
      (\mathit{step} = \mathsf{none} \land z~s = \mathsf{some}~b)\ \lor
      (\exists\ x\ b',\ \mathit{step} = \mathsf{some}~(x, b') \land f~x~b'~s = \mathsf{some}~b) \\
      &g'~b~s = \\
      &\quad g~b~s \bind \lambda\ \mathit{mstep} \Rightarrow \\
      &\quad\mathsf{match}~\mathit{mstep}~\mathsf{with}
      \mid \mathsf{none} \Rightarrow \gen{pure}~\mathsf{none}
      \mid \mathsf{some}~(x, b') \Rightarrow \gen{pure}~(\mathsf{some}~(x, (b', \mathit{st}~x~s)))
    \end{aligned}
  \end{gathered}
\]
\setlength{\jot}{3pt}%
}%
\noindent This rule is not pretty, but its basic operation is exactly the same
as \textsc{S-List-Unfold'}. It just allows for more flexible handling of hidden
invariants (see $\mathsf{isAllTwo}$) and more interesting
state passing.

\begin{wrapfigure}[8]{r}{.5\textwidth}
  \vspace{-5px}
  \centering
  \begin{minipage}{.5\textwidth}
{\small
\begin{lstlisting}
def genSorted : Gen (List @$\mathbb{N}$@) :=
  List.unfold (fun (lo, ()) =>
      pick (pure none)
           (pick (greaterThan lo) (pure lo) >>=
            fun x => pure (some (x, (x, ())))))
    (0, ())
\end{lstlisting}
}%
  \end{minipage}
\caption{A generator of sorted lists.}\label{fig:genSorted}
\end{wrapfigure}

If we rewrite $\mathsf{isSorted}$ from the beginning of this section using
$\mathsf{List.accuOpt}$ instead of $\mathsf{List.accu}$,
we can use
{\sc S-List-Unfold} to derive the generator in \autoref{fig:genSorted},
which behaves the same as one an expert user might write. At each step,
it either ends the list or generates a new value $x$ that is greater
than or equal to $\mathit{lo}$, puts it in the list, and continues with
$\mathit{lo} = x$.

    \subsection{Tupling Predicates}\label{sec:folds:tupling}

    The {\sc S-List-Unfold} rule works for predicates that are written as a single
    pass over a data structure, but sometimes predicates have multiple independent
    constraints. For example,
    {\small
    \begin{center}
      \code{def isAllTwoLengthK (k :\ $\mathbb{N}$) (xs : List\ $\mathbb{N}$) := isAllTwo xs && isLengthK k xs}
    \end{center}
    }%
    \noindent combines two predicates that we have seen before into a single predicate.

    We have two options for handling such situations. The first is to draw from the
    literature on program calculation and apply a {\em tupling}
    transformation~\cite{pettorossiUseTuplingStrategy1993,birdUsingCircularPrograms1984}
    to combine the conjuncts. These transformations were originally designed for
    optimizing functional
    programs, but they are a natural fit for this problem. Another is to
    adapt the ``merging'' procedure for inductive relations described by
    \citet{prinzMergingInductiveRelations2023}. Indeed it turns out that, in the case of
    predicates written with $\mathsf{List.accuOpt}$, these concepts coincide!

    We introduce a transformation $\mathsf{tupleAccuOpt}$ that takes two predicates
    $P$ and $Q$, each expressed with $\mathsf{accuOpt}$, and combines them into a
    single predicate---also expressed with $\mathsf{accuOpt}$---that computes
    $P~\mathit{xs} \land Q~\mathit{xs}$. The
    definition is surprisingly straightforward---the state and collector arguments
    are simply combined in a tuple and computed in parallel---so we do not show
    it here. This transformation means that our synthesis approach automatically
    benefits from the merging optimizations that users can apply manually
    in QuickChick~\cite{prinzMergingInductiveRelations2023}.

    Tupling is not the only useful program
    transformation that has been proposed for
    recursive programs in the program calculation literature. If we find
    that other transformations (e.g., fusion) are
    useful for predicates that users care about, we could easily extend
    our pipeline with them.

    \subsection{Putting it All Together}\label{sec:folds:ergonomics}

    The machinery described in the previous subsections is assembled as
    follows into
    a standardized workflow for transforming predicates on recursive data
    structures into a normal form
    (see \autoref{fig:fold-workflow}).
    \begin{enumerate}
      \item Normalize the predicate to the form
        $\lambda\ \mathit{xs} \Rightarrow F_1~\mathit{xs} = b_1 \land \cdots
        \land F_n~\mathit{xs} = b_n$.
      \item Rewrite each $F$ as a fold. Concretely, search for $z$ and
        $f$ satisfying the equations $F~[] = z$ and $F~(x :: xs) = f~x~(F~xs)$.
        If these equations are satisfied, then $F$ can be rewritten as
        $\mathsf{List.fold}~f~z$. This is sometimes referred to as the ``universal
        property of fold''~\cite{malcolm1990algebraic,huttonTutorialUniversalityExpressiveness1999}.
      \item Rewrite each fold as an optional accumulation based on its return type.
        For example, $\mathsf{isAllTwo}$, which collects an always-true boolean, would
        be transformed differently from $\mathsf{isSorted}$, which collects a higher-order
        function.
      \item Tuple the $n$ different branches of the predicate together.
    \end{enumerate}
    After normalization, we apply {\sc S-List-Unfold}
    to the final accumulation and
    obtain a generator by recursively synthesizing the step function.  This workflow
    is automated through tactics in Lean, which we discuss in the next
    section.

    Everything in this section has been phrased in terms of the $\mathsf{List}$ data
    type, but there is actually nothing in the workflow that is specific to
    lists. Any recursive data type composed of products and sums admits
    operations analogous to $\mathsf{List.fold}$, $\mathsf{List.accuOpt}$,
    etc.
    This means that the
    pipeline from \autoref{fig:fold-workflow} directly generalizes to a wide
    range of recursive data structures.

    The case studies in \autoref{sec:evaluation} required implementing this
    pipeline (plus other utilities, e.g., for case splitting) for five data
    structures: lists, binary trees, STLC types, STLC terms, and stacks
    (from~\cite{hritcuTestingNoninterferenceQuickly2013}). The
    implementation process is quite
    mechanical, as the definitions follow the structure of the data type and its
    constructors, and it should be straightforward to automate
    it, either via
    meta-programming or using ``quotients of polynomial functors''
    (QPFs)~\cite{avigadDataTypesQuotients2019}. We leave this engineering
    for future work.

\section{\rawimpname{}: Synthesizing Generators in Lean}\label{sec:implementation}

In this section we describe some implementation details of \impname.
  We begin with an overview
(\autoref{sec:implementation:overview}), then describe the
synthesis algorithm in detail (\autoref{sec:implementation:algorithm}).

\subsection{Overview}\label{sec:implementation:overview}

\impname{} is packaged as a library for the Lean theorem
prover. Using Lean's powerful meta-programming capabilities, everything can
be implemented as standard Lean code, and using it requires only importing the
library.
Then,
starting with the BST validity predicate from
\autoref{fig:bst-predicate}, a user can invoke \impname's
\code{generator_search} tactic
to synthesize a generator for BSTs:
\begin{lstlisting}
  def genBST (lo hi : Nat) : Gen (Tree Nat) := by
    generator_search (fun t => isBST t (lo, hi) = true)
\end{lstlisting}
When the Lean compiler builds the file, it runs \code{generator_search}
to synthesize an appropriate generator---in this case, \code{genBST} from
\autoref{fig:bst-unfold}.

While there are no proofs visible in this workflow, they exist under
the hood. The \code{generator_search} tactic proves that the
generator it synthesizes is correct with respect to the provided
predicate and assume-free. Alternative versions of
\code{generator_search} give users access to those proofs if
needed.

\subsection{The Synthesis Algorithm}\label{sec:implementation:algorithm}

We can think of ``\code{generator_search P}'' as calling three tactics in order:
\code{synthesize}, \code{optimize}, and \code{prove_assume_free}.
We now describe each of these steps in more detail.

\mypara{Step 1: Synthesize}
The \code{synthesize} tactic solves a goal of type
\code{CorrectGen P} by applying the rules in
\autoref{sec:theory:basic}. The procedure uses Aesop, a tactic in
Lean that performs best-first proof
search~\cite{limpergAesopWhiteBoxBestFirst2023}. Aesop takes a large set of
Lean tactics and applies them in a loop; when all tactics fail to solve a
particular sub-goal%
, the
search backtracks to try a different route.
Aesop has a timeout, which the user can change.
Relying on Aesop in this way turns
out to be extremely effective in terms of  both the results for our case studies and
ease of implementation.
The rules we provide to Aesop mirror the ones described in
\autoref{sec:theory}; they are given in full in
\autoref{appendix:concrete-synthesis-rules}.

\mypara{Step 2: Optimize}
The \code{optimize} tactic applies the optimization rules
from
\autoref{sec:theory:optimization}. It is implemented as a meta-level function,
operating directly on the AST of the generator. This makes it easy to write
rules like rule (4) from \autoref{sec:theory:optimization}, which
matches on the body of a lambda abstraction. Being
written at the meta level means that we cannot prove once and for all that
optimization is correct in Lean (though it is straightforward to prove
on paper).
Instead, we use proof automation to show that each optimized generator is
equivalent to its unoptimized counterpart.

\mypara{Step 3: (Optionally) Prove Assume-Free}
After optimization, we use straightforward proof automation to check whether the
synthesized generator makes nontrivial use of the $\gen{assume}$ constructor.
All but three generators we synthesize in our evaluation are assume-free, and
the automation easily proves that fact; we emit a warning if the assume-free
check fails.
We say that this step is optional because \impname{} does not categorically
reject generators that contain assumes. For example, two benchmarks that we
borrow from \citet{prinzMergingInductiveRelations2023}, AVL trees and red-black
trees, make nontrivial use of the $\gen{assume}$ constructor. We discuss this
more in \autoref{sec:evaluation:stlc}.

\mypara{Step 4: Render to the User}
At this point, the user has a choice. If they think their predicate may
change over the course of development, or if they just want to keep the
codebase simple, they can choose to leave the call to
\code{generator_search} as the definition of their generator. Lean
will try to cache the generator when possible; otherwise it will
re-synthesize the generator when reloading the file.
Users may also want to {\em render} the synthesized generator as
a concrete program that they can edit. They might, for example,
want to add weights to bias the distribution or add size
bounds to ensure generated values do not get too big.  (This may make the
  generator incomplete, so we do not do it during synthesis, but an
  expert user may
decide the incompleteness is worth it on a case-by-case basis.)
The \code{generator_search?} tactic invokes the same
synthesis procedure and then provides the user with a ``try this''
widget in
their editor; clicking pastes the text of the generator into
their file.

\mypara{Step 5: Interpret and Run the Generator}
As discussed in \autoref{sec:theory}, the generators we
synthesize are data structures, not programs, so they cannot be run directly.
A {\em sampling interpreter} gives meaning to generators as maps
from random seeds
to values, mirroring \autoref{def:support}.
In order to be consistent with a generator's support, the sampling
interpretation needs to handle backtracking and non-termination carefully.
Specifically, it re-samples values in the case of failure (e.g., from
$\gen{assume}$), and re-tries functions wrapped by $\gen{indexed}$
with exponentially increasing fuel if they run out.

  \section{Evaluation}\label{sec:evaluation}

  To evaluate \impname, we first examine its performance,
  comparing it with two state of the art competitors.
  We also qualitatively compare our generators to handwritten
  ones from the literature.
  We address the following research questions:
  \begin{itemize}
    \item[\bf RQ1] How does \impname compare to Cobb, a recent
    approach that uses program synthesis to repair incomplete PBT generators?
    \item[\bf RQ2] How does \impname compare to QuickChick, the standard tool
    for deriving generators from inductively defined specifications in Rocq?
    \item[\bf RQ3] How do the generators synthesized by \impname{} compare to
      ones that expert users write?
  \end{itemize}

  \mypara{Benchmarks}
  We use several sets of benchmark to answer these questions:
  \begin{itemize}[leftmargin=15pt]
    \item \mainbench: A set of 32 predicates demonstrating specific
      aspects of \impname{}'s synthesis algorithm, including low-level
      examples from
      the text above and others drawn from the PBT
      literature~\cite{lampropoulosGeneratingGoodGenerators2017,
      prinzMergingInductiveRelations2023, lafontaineWeveGotYou2025}.
      A detailed table of the benchmarks is in \autoref{appendix:extended-benchmarks}.
    \item \cobbbench:
      To compare to the state of the art in RQ1, we used versions of
      the benchmarks in \mainbench in the input format for
      Cobb~\cite{lafontaineWeveGotYou2025}, a synthesis-guided repair tool for generators.
      For benchmarks that were not drawn from Cobb's evaluation,
      we needed to create sketches of generators to repair.
      Following Cobb's existing benchmarks,
      we turned each target generator into two sketches:
      a \emph{minimal sketch} that left only the control flow of the generator with holes,
      and a \emph{maximal sketch} that only removed one element from the generator.
      (Of the possible elements, we removed the one that allowed Cobb to
      perform best.)
    \item \textsc{Inductive}: To test RQ2,
      we used versions of the same 32 benchmarks defined as Rocq inductive predicates,
      so that they are a fit for QuickChick~\cite{lampropoulosGeneratingGoodGenerators2017}.
      Many of the benchmarks (such as the tree- and STLC-based ones) already had inductive
      variants, as they were themselves originally adapted from
      QuickChick's benchmark suite~\cite{lafontaineWeveGotYou2025};
      the rest were straightforward to translate manually.
  \end{itemize}
For RQ3, we implemented a selection of generators from \textsc{Main} by hand,
and below we qualitatively compare those generators with the corresponding
synthesized ones.

  \mypara{Experimental setup}
  All experiments were performed on an M1 MacBook Pro with 8 cores and 16\,GB of memory using Lean
  v4.21.0,
  with a timeout of 900 seconds.
  Time measurements of \impname are averaged over 30 runs and measured in seconds.
  Cobb was run from a local installation build following
  the instructions provided by the Cobb artifact~\cite{cobbArtifact}.
  The internal cost used by Cobb in lieu of a timeout was set to 50,000,000,
  cancelling out cases where Cobb ran out of ``cost'' before our timeout.

  \subsection{RQ1: Comparison with Cobb}\label{sec:evaluation:benchmarks}

  \begin{figure}[t]
  \centering
    \includegraphics[height=131pt]{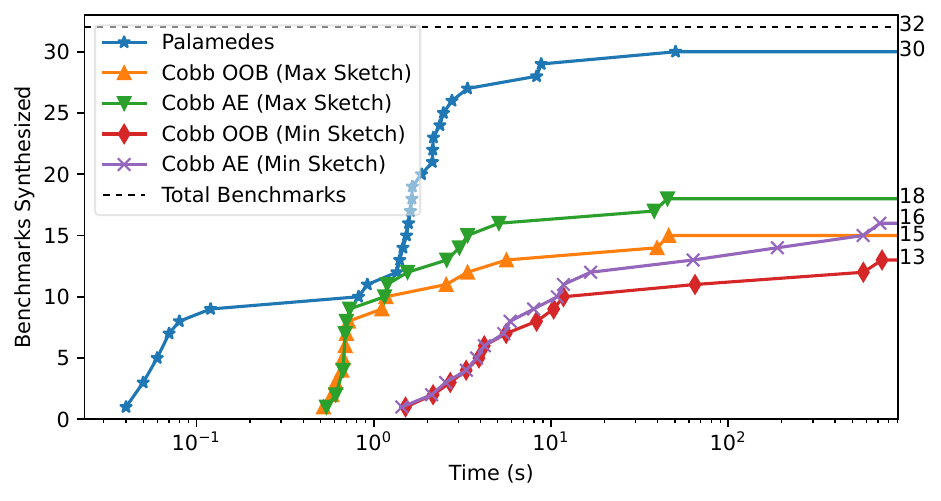}\quad
    \includegraphics[height=131pt]{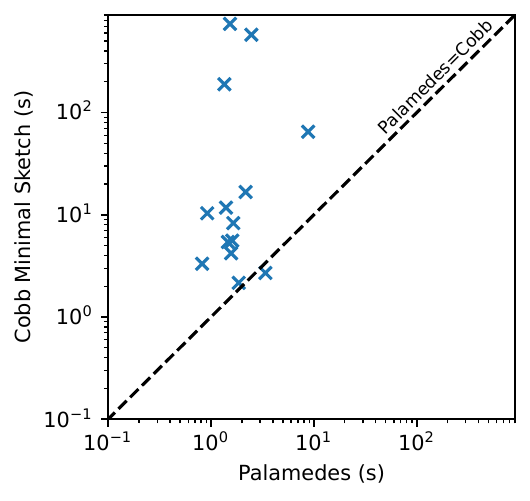}
    \caption{Comparing \impname and Cobb. Left: Number of generators
      synthesized by \impname{} and Cobb (OOB = ``Out of the Box'', AE = ``Additional Engineering''). Right: Per-benchmark comparison for the 16 benchmarks
      synthesized with both sketches (above the line means \impname is faster).
    Time axes are logarithmic.
    }\label{fig:cobb-comparison}
\Description{Plots of Cobb performance vs. Palamedes performance.}
  \end{figure}

  We ran each of the \mainbench benchmarks in \impname and their \cobbbench counterparts in Cobb
  and measured the time until a generator was found by each system.
  The results are shown in \autoref{fig:cobb-comparison}.
  Synthesis times for \impname{}
  range from around 40ms
  to 50s on \code{isRBT}, which has multiple nested case splits.
  \impname finds generators for 80\% of the \mainbench benchmarks in under 3
  seconds.

The two benchmarks that \impname fails to solve are \code{isUnique} and \code{hasDuplicates}
from Cobb's benchmark set. These examples have a large gap between their
recursive forms and their $\mathsf{accuOpt}$ form, and \impname currently cannot
automate that conversion---%
Aesop fails to make progress and gives up in under 2s for both.
Cobb has the benefit of a sketch to help scaffold
these tricky cases.

  From maximal sketches (the generator with one hole)
  Cobb's fastest benchmark, \code{isSortedBetween}, takes just over half a second.
  With minimal sketches, i.e., only the generator's control flow,
  Cobb solves its fastest benchmark, \code{isUnique}, in just over 1.5 seconds,
  and it takes 12 minutes to synthesize a generator for \code{isBST}.
  In total, Cobb solves 13 of the 32 benchmarks with a minimal sketch and 15
  with a maximal sketch out of the box. Our team spent a collective 20 person-hours
  attempting to adapt Cobb to the 19 missing benchmarks, and we were able to get
  another 3 working with both maximal and minimal sketches. Of the other 16,
  9 were very simple (e.g., $\mathbf{v} =
  2$); Cobb outputs ``nothing to repair,'' even though the generator is incomplete.
  We are certain Cobb can handle these in theory; that was likely a bug in our
  setup or the implementation.
  The other 7 were more complicated (e.g., the Palamedes version of \code{isRBT},
  which is stronger than the one in the Cobb paper). One of these
  (\code{isTrue}) cannot be encoded because the current implementation only
  supports integer lists; the rest might be implementable with more effort.
  
  Cobb was faster than \impname on one benchmark---\code{isLengthKAllTwos}---by 
  680ms (25\%). Here, \impname needs to find the case split on \code{length},
  whereas Cobb gets this decision as input in the provided sketch.

  Overall, \impname outperforms Cobb while requiring less information
  to produce a generator: only the predicate, with no additional sketch.
  We find that \textbf{\impname is more expressive than Cobb and
  finds generators faster, despite its search needing to solve
  a harder problem.}

\subsection{RQ2: Comparison with QuickChick}

\impname{} and Cobb use similar techniques to solve similar
problems.  By contrast, a direct comparison to the next-closest
state-of-the-art tool, QuickChick, is harder.
QuickChick does not search for generators; instead it works more like a
compiler, producing a generator almost instantly by walking over a predicate
in a single pass.
The trade-off is that QuickChick's predicates must be expressed as (slightly
restricted) inductive
datatypes in Rocq, rather than as functions.

To examine the relationship between the structure of the
predicate and the quality of the derived generator, we
ported the \mainbench benchmarks to QuickChick
and characterized the effort needed to achieve non-backtracking generators. The
results appear in detail in Appendix~\ref{app:QC}; we summarize and
discuss them here.

All 32 of the benchmarks needed to be in inductive
form: for some, inductive versions were already present in QuickChick's examples
as they were adapted from there; the rest were converted manually. Of the latter,
one ($\exists a,\ a = 3 \lor \mathbf{v} = a + 1$) could not be straightforwardly
ported as QuickChick does not handle existential quantifiers.
To effectively handle conjunctions of predicates
constraining the same variable, QuickChick requires users to
manually {\em merge} the conjoined
predicates~\cite{prinzMergingInductiveRelations2023}.
Ten of the benchmarks,
e.g., \code{AVL}, required manually invoking the merging procedure.
Finally, one benchmark, STLC's \code{isWellTyped}, also required a
specific constraint ordering in its definition for effective
generation, which is not a concern for
\impname.
Moreover, any predicates expressed naturally in a higher-order style
(e.g. using folds) would need to be manually translated to variants
using direct recursion.

In general, QuickChick's automation works extremely quickly when it works, but
it may require some careful setup from the user (including translation to an
inductive data type) before everything falls into place. By contrast, \impname
can search for solutions that do not follow the precise structure of the
predicate. This requires less user setup, especially when predicates are defined as
functions, but it is slower and less predictable.
Ultimately, we find that \impname \textbf{is a useful complement to QuickChick,
applying more directly to the common case where predicates are expressed as
functions, but with trade-offs in terms of performance.}

\subsection{RQ3: Synthesized vs. Handwritten Generators}\label{sec:evaluation:stlc}

Finally, we explore how the generators that \impname{} produces
compare to ones that expert users might write by hand.

First, we consider our synthesized generator for well-typed STLC terms,

shown in full in
\autoref{appendix:synthesized-genWellTyped}. The generator we synthesize
naturally implements a method for generating well-typed terms popularized by
prior work~\cite{palkaTestingOptimisingCompiler2011}, in which the generator
first generates a random type and then generates a random term of that type.

For comparison, we provide a handwritten STLC generator that the authors wrote
in \autoref{appendix:manual-stlc}. The handwritten generator reduces a bit of
code repetition and eta-reduces a match expression, but the asymptotic
performance of the generators would be the same---neither would backtrack, and
both would contain the same basic clauses.
Fundamentally, both of these generators implement the same generation algorithm
with the same time complexity.

Next, we consider our synthesized generator for AVL trees. \impname's AVL tree
generator appears in \autoref{appendix:synthesized-genAVL}. As we mentioned
briefly in \autoref{sec:implementation:algorithm}, this generator is one of the
few that we cannot remove all $\gen{assume}$s from. The generator we synthesize
is still quite efficient, and
the AVL tree
generator that \citet{prinzMergingInductiveRelations2023} present in their paper
on merging inductive relations also backtracks in the same way, but nevertheless
our generator is not perfect.

There are two common ways to generate AVL trees without any backtracking, which
we show in \autoref{appendix:manual-genAVL}. The first is a particularly
high-effort option that requires careful insight and analysis to get right. In
practice, it is uncommon for real users to go to this effort. More common is an
approach that generates an AVL tree by first generating a list of values and
then \code{insert}ing those values one by one into an empty tree. We could
imagine \impname looking for a generator like that someday, since it is
technically possible to express with our generator language, but we leave this
to future work.

\autoref{appendix:side-by-side} shows four more generators that \impname{}
synthesized, side-by-side with ones that we wrote manually. For all of these,
there were no functional differences between the generators we wrote and the
synthesized ones.

To summarize, we find that \textbf{\impname often implements the same
fundamental generation algorithm as handwritten versions (i.e., with the same
time complexity), but this does not necessarily mean that synthesized generators
are as fast to run (e.g., due to synthesis artifacts or tuning issues).}

  \section{Limitations}\label{sec:limitations}
  Our approach has some limitations.
  First, as with any synthesis algorithm, \impname is necessarily partial; there are many Lean
  predicates for which \impname{} does not successfully find a
  generator. (We discussed some of these in
  \autoref{sec:evaluation:benchmarks}.)
  A core reason for this is our handling of recursive predicates; by handling
  only predicates that can be expressed in terms of \code{accuOpt}, we limit
  ourselves to only first- and second-order primitive recursive programs.
  Synthesis may also fail due to missing base generators: for example,
  uses of $\gen{assume}$ need to be explicitly included for synthesis
  (e.g., in the definitions of {\sc S-Choose} and {\sc S-Elem}), so generators that
  require backtracking (e.g., the AVL example) may require manual extension of
  the library.
  Luckily, \impname{} is indeed
  extensible---users can add new proof automation, library generators, and
  synthesis rules---so its partiality can be mitigated over time. In
  \autoref{sec:future}, we discuss plans to extend the set of predicates that
  \impname{} can handle.

  Second, as we mention in \autoref{sec:intro}, this paper works with generators
  as nondeterministic programs, rather than as true probability distributions.
  We discuss how we hope to \emph{tune} our synthesized generators, optimizing
  their distributions as well, in \autoref{sec:future}.

  Finally, \impname{} does not
  incorporate {\em sizes}, a classic PBT generator feature.
  QuickCheck generators~\cite{claessenQuickCheckLightweightTool2000} have built-in
  ways to vary the size of generated values over the course of generation, which
  we do not include in our $\Gen$ type. Our
  approach is compatible with sized generation---prior  work~\cite{paraskevopoulouFoundationalPropertyBasedTesting2015}
  gives a semantics for internal sizing that plays well with our definitions---but
  we do not yet know how the synthesizer should use
  sizes. Currently, then,
  there are two ways to control the sizes of generated values in
  \impname:
\begin{enumerate*}
\item Users
  can augment their specifications with explicit size constraints. E.g.,
  \code{isLengthKAllTwos}
has an argument \code{k} to control the length of the generated list; the tupling
  transformation means this can be done by simply conjoining an extra predicate.
\item Users can manually modify their generators after synthesis to add
  size control.
\end{enumerate*}

\section{Related Work}\label{sec:related}

\mypara{The Constrained Random Generation Problem}
As discussed in \autoref{sec:intro}, many approaches to the constrained random
generation problem for testing have been proposed over the years. In rough
order of publication:
  \citet{deweyAutomatedDataStructure2015} propose input generation via
  constrained logic programming (CLP). This can handle some realistic examples
  like ``red-black trees that are guaranteed to rebalance when a particular
  input is inserted'' that would likely require some extra infrastructure in
  Palamedes.
  \citet{seidelTypeTargetedTesting2015} search for valid inputs via SMT
  in a refinement type system; their technique asks a solver for an example of
  a valid input, then an example of a valid input that isn't the first they
  were given, and so on.
  \citet{claessenGeneratingConstrainedRandom2015} use laziness in
  Haskell to find valid inputs, filtering bad choices as early as possible
  provided that the precondition is can be executed sufficiently lazily.
  \citet{reddyQuicklyGeneratingDiverse2020} guide generation via
  reinforcement learning, maximizing the chance that the next generated input
  is both valid and different from past inputs.
  \citet{goldsteinParsingRandomness2022} rely on Brzozowski derivatives
  to focus in on particular parts of the input space that might be more dense
  with valid values.
  \citet{steinhofelInputInvariants2022} again use SMT for generation,
  but this time in a stand-alone system that layers the solver's feedback on
  top of grammar-based generation.
  \citet{xiaFuzz4AllUniversalFuzzing2024} call a large language model to
  search for inputs that satisfy a function's preconditions.

These related approaches are all exciting and useful in various cases, but they
solve a fundamentally different problem from the one solved by \impname:
they actively guide generation towards valid values {\em during testing}, rather than
searching for a correct generator ahead of time.
In particular, all of these algorithms do potentially exponential work when
generating each input, while an assume-free generator produced by Palamedes
often does work linear in the size of the value being generated.
In cases where a generator is written once and then run many times, which is
common in practice~\cite{goldsteinPropertyBasedTestingPractice2024}, paying the
cost of search up-front is a good investment.

\mypara{The Constrained Generator Synthesis Problem}
As we discuss in detail in \autoref{sec:evaluation}, both
QuickChick~\cite{lampropoulosGeneratingGoodGenerators2017,
paraskevopoulouComputingCorrectlyInductive2022} and
Cobb~\cite{lafontaineWeveGotYou2025} address the constrained generator synthesis
problem in as we define it.

There are two other related tools that we did not get a chance to evaluate
against, since both are predecessors of and inspiration for the generator
automation in QuickChick.
First, Luck~\cite{lampropoulosBeginnerLuckLanguage2017} is a standalone
language for writing predicates (with some light annotations) from which
reasonably efficient generators could be derived. Luck's predicate language is
almost completely subsumed by QuickChick's handling of inductive relations.
Second, before the work in Rocq there was work on deriving generators in
Isabelle~\cite{Bulwahn12smartgen,Bulwahn12}. The work in Isabelle is different
from QuickChick in a few ways, and Isabelle can  handle some predicates
expressed as functions (rather than inductive relations). That said, the approach still
fundamentally follows the structure of the predicate to derive a generator,
rather than implementing a more flexible search like \impname does. We would
like to compare our approach to both Luck and Isabelle in the future.

\mypara{Deductive Program Synthesis}
Deductive program synthesis takes a specification in some logic
and searches for a proof that the specification holds;
proofs found by the search can be read as correct-by-construction programs.
  The simplest version of deductive synthesis is {\em type-directed synthesis}~\cite{10.1145/2491956.2462192,oseraTypeandexampledirectedProgramSynthesis2015,oseraConstraintbasedTypedirectedProgram2019,10.1145/2837614.2837629},
  where proofs are by application of typing rules. This has been
  extended to expressive type systems like refinement
  types~\cite{polikarpovaProgramSynthesisPolymorphic2016} and semantic
  types~\cite{10.1145/3519939.3523450}.
(Cobb~\cite{lafontaineWeveGotYou2025} also performs type-directed synthesis to generate repairs,
but uses a \emph{bottom-up} approach instead of a deductive one.)
  Several systems extend this approach to other
  logics~\cite{10.1145/2509136.2509555,10.1145/2676726.2677006,10.1145/3133889},
  including separation
  logic~\cite{itzhakyCyclicProgramSynthesis2021,polikarpovaStructuringSynthesisHeapmanipulating2019,10.1145/3591278}.

  \impname{} builds on the extensive body of work on deductive synthesis,
  particularly on
  SuSLik~\cite{polikarpovaStructuringSynthesisHeapmanipulating2019} and its
  followup work~\cite{10.1145/3591278,itzhakyCyclicProgramSynthesis2021,10.1145/3473589}.
  SuSLik synthesizes imperative programs by iteratively applying a proof rule
  for the next statement and leaving the remaining obligation to specify the rest of the program;
  we take inspiration from this process in how \impname handles $\gen{bind}$s.
  Our use of anamorphisms (unfolds) to build generators is closely related to
  \citet{hongRecursiveProgramSynthesis2024}'s use of {\em
    paramorphisms} for synthesizing
  recursive algorithms.
  However,
  \impname's target domain has not been explored by
  prior work, and its implementation differs in key ways. Unlike prior
  work, \impname{} relies on Lean for many aspects of proof search;
  e.g., avoiding explicit reasoning about termination and solving auxiliary
  lemmas with Aesop~\cite{limpergAesopWhiteBoxBestFirst2023}.
  And, excepting Fiat~\cite{10.1145/2676726.2677006} and SuSLik~\cite{10.1145/3473589}, prior
  work does not produce mechanized correctness proofs.  \impname's deduction
  rules are proved correct in the Lean theorem prover, and those proofs are
  combined to prove the final generator is correct,
  strengthening the guarantee that resulting programs are correct by
  construction.

  \section{Conclusion and Future Work}\label{sec:future}

\impname addresses the constrained generator synthesis problem, offering an
  algorithm for synthesizing generators that are correct with respect
  to a predicate, including generators for recursive data structures like lists
  and trees. Our approach combines prior work on deductive synthesis, functional
  programming, theorem provers, and more into a technique that has the potential to
  significantly advance PBT automation.
  We close with some ideas for future work.

  \mypara{Tuning Generator Distributions}
  The generators produced by \impname{} are guaranteed to produce the right set of
  values, but they may sample from those values with a suboptimal {\em
  distribution}.
  For example, the predicate
  $a = 1 \lor a = 2 \lor a = 3 \lor a = 4$
  will yield the generator
  \begin{center}
    \code{pick (pure 1) (pick (pure 2) (pick (pure 3) (pure 4)))}
  \end{center}
  which (assuming \code{pick} chooses branches with equal probability) will produce
  \code{1} with probability $0.5$,
  \code{2} with probability $0.25$, and
  \code{3} and
  \code{4} with probability $0.125$.
  This bias towards \code{1} is probably not something the user
  wants.

  Luckily, there are ways to address this problem. As a simple solution, we could
  implement an optimization pass that re-associates right-heavy
  nested sequences of \code{pick}s to
  prefer more balanced trees.  Less na\"ively, recent work has shown that probabilistic
  programming languages like Loaded Dice can be used to
  automatically tune generators to user-specified
  distributions~\cite{tjoaTuningRandomGenerators2025}. We plan to implement a
  translation from our generators into Loaded Dice, giving users comprehensive
  control over generator distributions.

  \mypara{Correct Generators for Everyday Developers}
  Implementing \impname{} as a Lean library has myriad benefits, but it has one
  major downside: theorem provers are relatively inaccessible to everyday software
  developers, so the tool in its current form is unlikely to see broad adoption.
  However, we see a clear path towards impact in the software engineering
  industry, by using \impname{} as a back end for user-facing tools. As a
  first step in this direction, we plan to (1) embed a subset of Python's
  semantics in Lean, (2) compile Python predicates to that sub-language, (3)
  synthesize generators for those predicates, and (4) render the synthesized generators as
  Hypothesis~\cite{maciverHypothesisNewApproach2019} strategies.
  If successful, we hope to push this paradigm even further, using \impname{} as a
  backend for synthesizing generators for other languages and PBT
  frameworks.

  \mypara{Better Automation and Algorithmic Improvements}
  As we demonstrated in \autoref{sec:evaluation}, \impname{} is already flexible
  enough to synthesize a wide range of generators, but the algorithm may be
  further improved by ongoing enhancements to Lean's proof automation. For
  example, Lean now has tactics for automating proof search with
  SMT solvers~\cite{mohamedLeanSMTSMTTactic2025} and e-graph
  rewriting~\cite{rosselBridgingSyntaxSemantics2024}.
  LLM-based proof automation is also an active area of study~\cite{liSurveyDeepLearning2024,
    xinDeepSeekProverV15HarnessingProof2024,
    yangLeanDojoTheoremProving2023,
    firstBaldurWholeProofGeneration2023,
    thompsonRangoAdaptiveRetrievalAugmented2025,
  kasibatlaCobblestoneIterativeAutomation2024}, which may unlock more flexible
  approaches to proof generation.

  These approaches dovetail nicely with our current infrastructure---they could
  be used to implement more powerful versions of our predicate simplifiers, and
  some could even to replace Aesop entirely as the engine for the core synthesis
  loop. We expect \impname{} to increase in power over time, in concert with the
  Lean community's improvements in proof automation.

\subsection*{Acknowledgements}
The work by Harrison Goldstein was funded by the Victor Basili Postdoctoral
Fellowship from the University of Maryland.

The work by Hila Peleg was funded by the European Union (ERC, EXPLOSYN,
101117232). Views and opinions expressed are however those of the authors only
and do not necessarily reflect those of the European Union or the European
Research Council Executive Agency.  Neither the European Union nor the granting
authority can be held responsible for them.

The work by Cassia Torczon, Daniel Sainati, Leonidas Lampropoulos, and Benjamin
C. Pierce was funded by a number NSF grants and fellowships: \emph{SHF: Medium:
Usable Property-Based Testing, NSF \#2402449}; \emph{CISE: Graduate Fellowships
Grant, NSF \#2313998}; and \emph{CAREER: Fuzzing Formal Specifications, NSF
\#2145649}. Any opinions, findings, and conclusions or recommendations expressed
in this material are those of the authors and do not necessarily reflect the
views of the National Science Foundation.

We also thank Mike Dodds, Shachar Itzhaky, Nadia Polikarpova, and Ilya Sergey
for their feedback and input at various stages of this project.

\subsection*{Data Availability Statement}
The code artifact for this paper is available on Zenodo at
\url{https://doi.org/10.5281/zenodo.19073205}.
We are also developing Palamedes as an open source project at 
\url{https://github.com/hgoldstein95/palamedes-lean} (at the time of writing, it
is still under active development).

\bibliographystyle{ACM-Reference-Format}
\bibliography{references}

\newpage

  \appendix

  \section{Derivation of a Simple Generator}\label{appendix:basic-generator-derivation}
  \begin{figure}[ht]
    \raggedright
    \[
      \inferrule*[right=?]
      { }
      { \cdot \ \vdash\ {?} \ :\  \mathsf{Gen}~(\lambda\ a \Rightarrow a = 1 \lor a = 2) }
    \]

    \vspace{1mm}
    \fbox{\bf apply {\sc S-Pick}}
    \vspace{1mm}
    \[
      \inferrule*[right=S-Pick]
      {
        \inferrule*[right=?]
        { }
        { \cdot \ \vdash\ {?x} \ :\  \mathsf{Gen}~(\lambda\ a \Rightarrow a = 1) }
        \\
        \inferrule*[right=?]
        { }
        { \cdot \ \vdash\ {?y} \ :\  \mathsf{Gen}~(\lambda\ a \Rightarrow a = 2) }
      }
      { \cdot \ \vdash\ \gen{pick}~?x~?y \ :\  \mathsf{Gen}~(\lambda\ a \Rightarrow a = 1 \lor a = 2) }
    \]

    \vspace{1mm}
    \fbox{\bf apply {\sc S-Pure} on the left}
    \vspace{1mm}
    \[
      \inferrule*[right=S-Pick]
      {
        \inferrule*[right=S-Pure]
        { \inferrule*{ }{\cdot \ \vdash\ 1 \ :\  \mathbb{N}} }
        { \cdot \ \vdash\ \gen{pure}~1 \ :\  \mathsf{Gen}~(\lambda\ a \Rightarrow a = 1) }
        \\
        \inferrule*[right=?]
        { }
        { \cdot \ \vdash\ {?y} \ :\  \mathsf{Gen}~(\lambda\ a \Rightarrow a = 2) }
      }
      { \cdot \ \vdash\ \gen{pick}~(\gen{pure}~1)~?y \ :\  \mathsf{Gen}~(\lambda\ a \Rightarrow a = 1 \lor a = 2) }
    \]

    \vspace{1mm}
    \fbox{\bf apply {\sc S-Pure} on the right}
    \vspace{1mm}
    \[
      \inferrule*[right=S-Pick]
      {
        \inferrule*[right=S-Pure]
        { \inferrule*{ }{\cdot \ \vdash\ 1 \ :\  \mathbb{N}} }
        { \cdot \ \vdash\ \gen{pure}~1 \ :\  \mathsf{Gen}~(\lambda\ a \Rightarrow a = 1) }
        \\
        \inferrule*[right=S-Pure]
        { \inferrule*{ }{\cdot \ \vdash\ 2 \ :\  \mathbb{N}} }
        { \cdot \ \vdash\ \gen{pure}~2 \ :\  \mathsf{Gen}~(\lambda\ a \Rightarrow a = 2) }
      }
      { \cdot \ \vdash\ \gen{pick}~(\gen{pure}~1)~(\gen{pure}~2) \ :\  \mathsf{Gen}~(\lambda\ a \Rightarrow a = 1 \lor 1 = 2) }
    \]
  \Description{A derivation of a simple generator.}
  \end{figure}
  \newpage

  \section{Greater Than and Less Than}\label{appendix:greater-than-less-than}

  \[
    \mathsf{def}~\mathsf{greaterThan}~(n : \mathbb{N}) :=
    \mathsf{arbNat} \bind \lambda\ \mathit{lo} \Rightarrow \gen{pure}~(\mathit{lo} + 1 + n)
  \]
  \begin{lemma}[Greater Than Support]
    $a \in \llbracket \mathsf{greaterThan}~\mathit{lo} \rrbracket \iff a > \mathit{lo}$.
  \end{lemma}
  \[
    \inferrule*[right=S-GreaterThan]
    { }
    { \Gamma \ \vdash\ \mathsf{greaterThan}~\mathit{lo} \ :\  \mathsf{Gen}_\mathbb{N}~(\lambda a \Rightarrow a > \mathit{lo}) }
  \]

  \[
    \mathsf{def}~\mathsf{lessThan}~(\mathsf{hi} : \mathbb{N}) := \mathsf{choose}~0~(\mathsf{hi} - 1)
  \]
  \begin{lemma}[Less Than Support]
    If $0 < \mathit{hi}$, then
    $a \in \llbracket \mathsf{lessThan}~\mathit{hi} \rrbracket \iff a < \mathit{hi}$.
  \end{lemma}
  \[
    \inferrule*[right=S-LessThan]
    { \Gamma \ \vdash\ 0 < \mathit{hi} }
    { \Gamma \ \vdash\ \mathsf{lessThan}~\mathit{hi} \ :\  \mathsf{Gen}_\mathbb{N}~(\lambda a \Rightarrow a < \mathit{hi}) }
  \]
  \[
    \inferrule*[right=S-LessThanPartial]
    { }
    { \Gamma \ \vdash\ \gen{assume}~0 < \mathit{hi}~\gen{in}\mathsf{lessThan}~\mathit{hi} \ :\  \mathsf{Gen}_\mathbb{N}~(\lambda a \Rightarrow a < \mathit{hi}) }
  \]

  \newpage

\section{Synthesis Rules for Aesop}\label{appendix:concrete-synthesis-rules}

\begin{table}[H]
  \caption{
    The synthesis rules used to synthesize our core examples. Each rule has a
    precedence; rules with 100\% precedence are tried first and never
    backtracked. Other rules are tried in order of precedence, and higher-precedence
    branches of the proof are explored first. Rules containing \code{<T>} are
    replicated once for each of the recursive data structures we consider.
  }\label{table:rules}
  \begin{tabular}{|l|l|}
    \hline
    {\bf Rule} & {\bf Precedence} \\\hline
    \code{uncurry_intro} & 100\% \\
    \code{apply s_arbitrary_<T>} & 100\% \\
    \code{assumption} & 99\% \\
    \code{apply convert (by match_pure) s_pure} & 99\% \\
    \code{apply convert (by match_pick) (s_pick _ _)} & 99\% \\
    \code{apply convert (by match_bind 0) (s_bind _ _)} & 99\% \\
    \code{apply convert (by match_bind 1) (s_bind _ _)} & 99\% \\
    \code{apply convert (by match_greaterThan) s_greaterThan} & 99\% \\
    \code{apply convert (by match_lessThan) s_lessThan_partial} & 99\% \\
    \code{apply convert (by match_between) (s_between (by solve_between))} & 99\% \\
    \code{apply convert (by match_between) s_between_partial} & 99\% \\
    \code{apply convert (by match_elements) s_elements_partial} & 99\% \\
    \code{apply convert (by match_<T>_unfold) (<T>.s_unfold _)} & 99\% \\
    \code{split_cases 0 <T>.split} & 5\% \\
    \code{split_cases 1 <T>.split} & 5\% \\
    \code{split_cases 2 Nat.split} & 5\% \\
    \code{split_cases 3 Nat.split} & 5\% \\
    \hline
  \end{tabular}
\end{table}

  \newpage

  \section{Predicate Definitions}\label{appendix:predicate-definitions}

\begin{lstlisting}
def List.fold {@$\alpha$@ @$\beta$@: Type} (f : @$\alpha$@ -> @$\beta$@ -> @$\beta$@) (z : @$\beta$@) (xs : List @$\alpha$@) :=
  List.foldr f z xs
inductive Tree (@$\alpha$@ : Type) where
  | leaf : Tree @$\alpha$@
  | node : (l : Tree @$\alpha$@) -> (x : @$\alpha$@) -> (r : Tree @$\alpha$@) -> Tree @$\alpha$@
def Tree.fold
    {@$\alpha$@ @$\beta$@ : Type}
    (f : @$\beta$@ -> @$\alpha$@ -> @$\beta$@ -> @$\beta$@)
    (z : @$\beta$@)
    (t : Tree @$\alpha$@) :
    @$\beta$@ :=
  match t with
  | .leaf => z
  | .node l x r => f (Tree.fold f z l) x (Tree.fold f z r)
inductive Label where
  | low
  | high
inductive Atom where
  | atm (n : Nat) (l : Label)
inductive Stack where
  | mty
  | cons (a : Atom) (s : Stack)
  | ret_cons (pc : Atom) (s : Stack)
def Stack.fold
  {@$\alpha$@ : Type}
  (z : @$\alpha$@)
  (f_c : Atom -> @$\alpha$@ -> @$\alpha$@)
  (f_rc : Atom -> @$\alpha$@ -> @$\alpha$@)
  (s : Stack) : @$\alpha$@ :=
  match s with
  | .mty => z
  | .cons x s' => f_c x (Stack.fold z f_c f_rc s')
  | .ret_cons pc s' => f_rc pc (Stack.fold z f_c f_rc s')
inductive Ty : Type where
  | unit
  | arrow (@$\tau$@1 @$\tau$@2 : Ty)
  deriving BEq
inductive Term : Type where
  | unit
  | var (n : Nat)
  | abs (@$\tau$@ : Ty) (t : Term)
  | app (t1 t2 : Term)
def Term.fold
    {@$\alpha$@ : Type}
    (z : @$\alpha$@)
    (zn : Nat -> @$\alpha$@)
    (f_abs : Ty -> @$\alpha$@ -> @$\alpha$@)
    (f_app: @$\alpha$@ -> @$\alpha$@ -> @$\alpha$@)
    (t : Term) :
    @$\alpha$@ :=
  match t with
  | .unit => z
  | .var n => zn n
  | .abs @$\tau$@ t' => f_abs @$\tau$@ (Term.fold z zn f_abs f_app t')
  | .app t1 t2 =>
    f_app (Term.fold z zn f_abs f_app t1) (Term.fold z zn f_abs f_app t2)
def isAllTwos : List Nat -> Bool
  | [] => true
  | x :: xs => x = 2 && isAllTwos xs
def isEvenLen : List @$\alpha$@ -> Bool
  | [] => true
  | _ :: xs => !(isEvenLen xs)
def isAllTwosEvenLen (xs : List Nat) : Bool :=
  isAllTwos xs && isEvenLen xs
def isIncreasingByOneAux (xs : List Nat) (prev : Nat) : Bool :=
  match xs with
  | [] => true
  | x :: xs' => x == prev + 1 && isIncreasingByOneAux xs' x
def isIncreasingByOne (xs : List Nat) : Bool :=
  isIncreasingByOneAux xs 0
def isLengthKAllTwos (k : Nat) (xs : List Nat) : Bool :=
  xs.length == k && isAllTwos xs
def isSortedBetween : List Nat -> Nat @$\times$@ Nat -> Bool := fun xs (lo, hi) =>
  match xs with
  | [] => true
  | x :: xs' => (lo <= x && x <= hi) && isSortedBetween xs' (x, hi)
def isTrue : List @$\alpha$@ -> Bool
  | [] => true
  | x :: xs => (fun _ => true) x && isTrue xs
def hasDuplicatesAux (xs : List @$\alpha$@) (soFar : List @$\alpha$@) :=
  match xs with
  | [] => false
  | x :: xs' => List.elem x soFar || hasDuplicatesAux xs' (x :: soFar)
def hasDuplicates (xs : List @$\alpha$@) :=
  hasDuplicatesAux xs []
def isUniqueAux (xs : List @$\alpha$@) (soFar : List @$\alpha$@) :=
  match xs with
  | [] => false
  | x :: xs' => !List.elem x soFar && isUniqueAux xs' (x :: soFar)
def isUnique (xs : List @$\alpha$@) :=
  isUniqueAux xs []
def isAllEvens : List Nat @$\to$@ Bool
  | [] => true
  | x :: xs => x % 2 = 0 && isAllEvens xs
def isRRAux : Tree (Color @$\times$@ @$\alpha$@) @$\to$@ Bool @$\to$@ Bool := fun t isRedChild =>
 match t with
 | .leaf => true
 | .node l c r => if c.fst == .red then !isRedChild && isRRAux l true && isRRAux r true else isRRAux l false && isRRAux r false
def isRR : Tree (Color @$\times$@ @$\alpha$@) @$\to$@ Bool := fun t => isRRAux t false
def isBH : Tree (Color @$\times$@ @$\alpha$@) @$\to$@ Nat @$\to$@ Bool := fun t height =>
 match t with
 | .leaf => height == 0
 | .node l c r => if c.fst == .red then isBH l height && isBH r height else height >= 0 && isBH l (height - 1) && isBH r (height - 1)
def isRBT : Tree (Color @$\times$@ Nat) @$\to$@ Nat @$\to$@ Nat @$\to$@ Nat @$\to$@ Bool := fun t height lo hi =>
  isRR t = true && isBST t (lo, hi) = true && isBH t height = true
def isRBTNoBST : Tree (Color @$\times$@ Nat) @$\to$@ Nat @$\to$@ Bool := fun t height =>
  isRR t = true && isBH t height = true
def isAllTwosFold (xs : List Nat) : Bool :=
  List.fold (fun x b => x == 2 && b) true xs
def isAllTwosEvenLenFold (xs : List Nat) : Bool :=
  List.fold (fun x b => x == 2 && b) true xs = true
    @$\land$@ List.fold (fun _ b => !b) true xs
def isEvenLenFold (xs : List @$\alpha$@) : Bool :=
  List.fold (fun _ b => !b) true xs
def isIncreasingByOneFold (xs : List Nat) : Bool :=
  List.fold (fun x b prev => x == prev + 1 && b x) (fun _ => true) xs 0
def lengthFold (xs : List @$\alpha$@) : Nat :=
  List.fold (fun _ b => b + 1) 0 xs
def isLengthKAllTwosFold (k : Nat) (xs : List Nat) :=
  List.fold (fun _ b => b + 1) 0 xs = k
    @$\land$@ List.fold (fun x b => x == 2 && b) true xs
def isSortedBetweenFold (lo hi : Nat) (xs : List Nat) : Prop :=
  List.fold (fun x b s => decide (s <= x)
    && decide (x <= hi) && b x) (fun _ => true) xs lo
def isTrueFold (xs : List @$\alpha$@) : Bool :=
  List.fold (fun _ b => b) true xs
def isAllEvensFold (xs : List Nat) : Bool :=
  List.fold (fun x b => x % 2 == 0 && b) true xs
def isAllTwos : Tree Nat -> Bool
  | .leaf => true
  | .node l x r => x = 2 && isAllTwos l && isAllTwos r
def isBST : Tree Nat -> (Nat @$\times$@ Nat) -> Bool := fun t @$\langle$@lo, hi@$\rangle$@ =>
  match t with
  | .leaf => true
  | .node l x r =>
    (lo <= x && x <= hi) &&
    isBST l @$\langle$@lo, x - 1@$\rangle$@ &&
    isBST r @$\langle$@x + 1, hi@$\rangle$@
def isComplete : Tree @$\alpha$@ -> Nat -> Bool := fun t n =>
  match t with
  | .leaf => n == 0
  | .node l _ r =>
    n > 0 &&
    isComplete l (n - 1) &&
    isComplete r (n - 1)
def isMaxDepth (t : Tree @$\alpha$@) (n : Nat) : Bool :=
    match t with
    | .leaf => true
    | .node l _ r =>
      n > 0 &&
      isMaxDepth l (n - 1) &&
      isMaxDepth r (n - 1)
def isIncreasingByOneAux (t : Tree Nat) (prev : Nat) : Bool :=
  match t with
  | .leaf => true
  | .node l x r =>
    x == prev + 1 &&
    isIncreasingByOneAux l x &&
    isIncreasingByOneAux r x
def isIncreasingByOne (t : Tree Nat) : Bool :=
  isIncreasingByOneAux t 0
def isNonempty : Tree @$\alpha$@ -> Bool
  | .leaf => false
  | .node l _ r => true && isNonempty l && isNonempty r
def isAllTwosFold (t : Tree Nat) : Bool :=
  Tree.fold (fun bl x br => x == 2 && bl && br) true t
def isBSTFold (lo hi : Nat) (t : Tree Nat) : Bool :=
  Tree.fold
        (fun bl x br s =>
          match s with
          | (sl, sr) =>
            (decide (sl <= x) &&
              decide (x <= sr)) &&
              bl (sl, x - 1) && br (x + 1, sr))
        (fun _ => true) t (lo, hi)
def isCompleteFold (n : Nat) (t : Tree Nat) : Bool :=
  Tree.fold (fun bl _ br s => decide (s > 0) &&
  bl (s - 1) &&
  br (s - 1)) (fun s => s == 0) t n
def isMaxDepthFold (t : Tree Nat) (n : Nat) : Bool :=
  Tree.fold (fun bl _ br s => decide (s > 0) &&
  bl (s - 1) &&
  br (s - 1)) (fun _ => true) t n
def isIncreasingByOneFold (t : Tree Nat) : Bool :=
  Tree.fold
    (fun bl x br prev => x == prev + 1 && bl x && br x)
    (fun _ => true) t 0
def isNonemptyFold (t : Tree @$\alpha$@) : Bool :=
  Tree.fold (fun _ _ _ => true) false t
def isBalanced : Tree Nat -> Nat -> Bool := fun t height =>
  match t with
  | .leaf => height <= 1
  | .node l _ r =>
    height > 0 &&
    isBalanced l (height - 1) &&
    isBalanced r (height - 1)
def isAVL (height lo hi : Nat) (t : Tree Nat) : Bool :=
  isBST t (lo, hi) && isBalanced t height
def isAVLFold (height lo hi : Nat) (t : Tree Nat) : Bool :=
  Tree.fold
      (fun bl x br bounds =>
        match bounds with
        | (sl, sr) => decide (sl <= x) && decide (x <= sr)
          && bl (sl, x - 1) && br (x + 1, sr))
      (fun _ => true) t (lo, hi) = true
    @$\land$@
    Tree.fold
      (fun bl _ br h => decide (h > 0) && bl (h - 1) && br (h - 1))
      (fun h => decide (h <= 1)) t height
def isRRFold (t : Tree (Color @$\times$@ @$\alpha$@)) : Bool :=
  Tree.fold
    (fun bl c br isRedChild => if c.fst == .red then !isRedChild && bl true && br true else bl false && br false)
    (fun _ => true)
    t
    false
def isBHFold (t : Tree (Color @$\times$@ @$\alpha$@)) (height : Nat) : Bool :=
  Tree.fold
    (fun bl c br h => if c.fst == .red then bl h && br h else h >= 0 && bl (h - 1) && br (h - 1))
    (fun h => h == 0)
    t
    height
def isRBTFold (height lo hi : Nat) (t : Tree (Color @$\times$@ Nat)) : Bool :=
  isBHFold t height = true && isRRFold t = true && isBSTFold t (lo, hi) = true
def isRBTNoBSTFold (height : Nat) (t : Tree (Color @$\times$@ Nat)) : Bool :=
  isBHFold t height = true && isRRFold t = true
def isGoodNat (n : Nat) : Bool :=
  n == 0 || n == 1
def isGoodAtom : Atom -> Bool
  | .atm n _ => isGoodNat n
def isGoodStackFold (s : Stack) (n : Nat) : Bool :=
  Stack.fold (fun s => s == 0)
    (fun x acc s => isGoodAtom x && acc (s - 1))
    (fun pc acc s => isGoodAtom pc && acc (s - 1)) s n
def isGoodStack (s : Stack) (n : Nat) : Bool :=
  match s with
  | .mty => n == 0
  | .cons x s' => (n > 0 && isGoodAtom x) && isGoodStack s' (n - 1)
  | .ret_cons pc s' => (n > 0 && isGoodAtom pc) && isGoodStack s' (n - 1)
def getType (t : Term) (@$\Gamma$@ : List Ty) : Option Ty :=
  match t with
  | .unit => pure .unit
  | .var n => @$\Gamma$@[n]?
  | .abs @$\tau$@ t => do
    let @$\tau$@' @$\leftarrow$@ getType t (@$\tau$@ :: @$\Gamma$@)
    pure (.arrow @$\tau$@ @$\tau$@')
  | .app t1 t2 => do
    let @$\tau$@1 @$\leftarrow$@ getType t1 @$\Gamma$@
    let @$\tau$@2 @$\leftarrow$@ getType t2 @$\Gamma$@
    match @$\tau$@1 with
    | .arrow @$\tau$@arg @$\tau$@res => do
      guard (@$\tau$@arg == @$\tau$@2)
      pure @$\tau$@res
    | .unit => failure
def isWellTyped (@$\Gamma$@ : List Ty) (t : Term) : Prop :=
  @$\exists$@ (@$\tau$@ : Ty), getType t @$\Gamma$@ = @$\tau$@
def isWellScoped : Term -> Nat -> Bool := fun t varCap =>
  match t with
  | .unit => true
  | .var n => n < varCap
  | .abs _ t => isWellScoped t (varCap + 1)
  | .app t1 t2 => isWellScoped t1 varCap && isWellScoped t2 varCap
def getTypeFold : Term -> List Ty -> Option Ty :=
  Term.fold
    (fun _ => pure .unit)
    (fun n @$\Gamma$@' => @$\Gamma$@'[n]?)
    (fun @$\tau$@1 b @$\Gamma$@' => do
      let @$\tau$@2 @$\leftarrow$@ b (@$\tau$@1 :: @$\Gamma$@')
      pure (.arrow @$\tau$@1 @$\tau$@2))
    (fun b1 b2 @$\Gamma$@' => do
      let @$\tau$@1 @$\leftarrow$@ b1 @$\Gamma$@'
      let @$\tau$@2 @$\leftarrow$@ b2 @$\Gamma$@'
      match @$\tau$@1 with
      | .arrow @$\tau$@arg @$\tau$@res => do
        guard (@$\tau$@arg == @$\tau$@2)
        pure @$\tau$@res
      | Ty.unit => failure)
def isWellTypedFold (@$\Gamma$@ : List Ty) (t : Term) : Prop :=
  @$\exists$@ @$\tau$@, getTypeFold t @$\Gamma$@ = some @$\tau$@
def isWellScopedFold (varCap : Nat) (t : Term) : Bool :=
  Term.fold
    (fun _ => true)
    (fun n s => s < n)
    (fun _ b s => b (s + 1))
    (fun b1 b2 s => b1 s && b2 s)
    t
    varCap
\end{lstlisting}

  \newpage

  \section{Extended Tables of Benchmarks}\label{appendix:extended-benchmarks}

  \subsection{\mainbench}

The following table lists the 32 benchmarks in the \mainbench benchmark set.

The value being generated is
  \code{$\mathbf{v}$}; other
  variables are universally quantified unless specified.  External definitions
  (e.g., \code{isBST}) are presented in \autoref{appendix:predicate-definitions}.
We list several measures of the complexity of the benchmark:
First, each type being generated lists the number of constructors for the ADT,
which indicates case splitting possibilities in the search.
Next, for each benchmark we also count the number of functions used in the specification,
e.g., \code{isAllTwosEvenLength} is specified using two auxiliary functions,
\code{isAllTwos} and \code{isEvenLength}.
Finally, we tally the number of recursive calls across all these functions.

\begin{table}[H]
    \begin{tabular}{|lr|l|rr|}
      \hline
                 & {\bf ADT} &                   & {\bf no. functions} & {\bf no. recursive} \\
      {\bf Type} & {\bf Cases} & {\bf Predicate} & {\bf in spec}       & {\bf calls (total)} \\\hline
      \multirow{9}{*}{\code{Nat}} & \multirow{9}{*}{2} & \code{$\mathbf{v}$ = 2} & 0  & 0 \\
      & & \code{2 =\ $\mathbf{v}$} & 0  & 0 \\
      & & \code{$\mathbf{v}$ = 2\ $\lor$\ $\mathbf{v}$ = 5} &0  & 0 \\
      & & \code{$\mathbf{v}$ = 2\ $\lor$\ $\mathbf{v}$ = 5\ $\land$\ True} &  0  & 0 \\
      & & \code{$\exists$ a, a = 3\ $\land$\ $\mathbf{v}$ = a + 1} &  0  & 0 \\
      & & \code{5 <=\ $\mathbf{v}$\ $\land$\ $\mathbf{v}$ <= 10} &  0  & 0 \\
      & & \code{$\mathbf{v}$ > 5} &  0  & 0 \\
      & & \code{lo <= $\mathbf{v}$\ $\land$\ $\mathbf{v}$ <= hi} & 0  & 0 \\
      & & \code{$\mathbf{v}$ = 0 $\lor$ lo <=\ $\mathbf{v}$\ $\land$\ $\mathbf{v}$\ <= hi} &  0  & 0 \\ \hline
      \multirow{11}{*}{\code{List Nat}} & \multirow{11}{*}{2} &
      \code{isAllTwos\ $\mathbf{v}$ = true} & 1 & 1 \\
      & & \code{isAllTwosEvenLen\ $\mathbf{v}$ = true} & 3 & 2 \\
      & & \code{isEvenLen\ $\mathbf{v}$ = true} & 1 & 1\\
      & & \code{isIncreasingByOne\ $\mathbf{v}$ = true} & 2 & 1\\
      & & \code{List.length\ $\mathbf{v}$ = k} & 0 & 0\\
      & & \code{isLengthKAllTwos k\ $\mathbf{v}$ = true} & 2 & 1\\
      & & \code{isSortedBetween\ $\mathbf{v}$ (lo, hi) = true} & 1 & 1 \\
      & & \code{isAllEvens\ $\mathbf{v}$ = true} & 1 & 1\\
      & & \code{isTrue\ $\mathbf{v}$ = true} & 1 & 1\\
      & & \code{isUnique\ $\mathbf{v}$} & 2 & 1\\
      & & \code{hasDuplicates\ $\mathbf{v}$} & 2 & 1\\
      \hline
      \multirow{9}{*}{\code{Tree Nat}} & \multirow{9}{*}{2} &
      \code{isAllTwos\ $\mathbf{v}$ = true} & 1 & 2\\
       & & \code{isBST\ $\mathbf{v}$ (lo, hi) = true} & 1 & 2\\
      & & \code{isComplete\ $\mathbf{v}$ n = true} & 1 & 2\\
      & & \code{isMaxDepth\ $\mathbf{v}$ n = true} & 1 & 2 \\
      & & \code{isIncreasingByOne\ $\mathbf{v}$ = true} & 2 & 2\\
      & & \code{isNonempty\ $\mathbf{v}$ = true} & 1 & 2\\
      & & \code{isAVL height lo hi\ $\mathbf{v}$ = true} & 3 & 4 \\
      & & \code{isRBTNoBST height\ $\mathbf{v}$ = true} & 4 & 8\\
      & & \code{isRBT height lo hi\ $\mathbf{v}$ = true} & 5 & 10\\
      \hline
      \code{Stack} & 3 &
      \code{isGoodStack\ $\mathbf{v}$ n = true} &  3 & 2\\
      \hline
      \multirow{2}{*}{\code{Term}} & \multirow{2}{*}{4} &
      \code{isWellScoped\ $\mathbf{v}$ 0 = true} & 1 & 3\\
      & & \code{isWellTyped\ $\Gamma$\ $\mathbf{v}$} & 2 & 3\\
      \hline
    \end{tabular}
\end{table}

\newpage

The following table lists the runtimes of \impname on the \mainbench benchmarks.
Means are presented with standard
deviations in parentheses.
Generators above the line are assume-free.
The four generators below the line are not:
AVL and RBT generators are known to need backtracking (e.g., the generator in \citet{prinzMergingInductiveRelations2023}).
The generator of a range is not assume-free because the $\gen{assume}$ introduced by the \textsc{S-ChoosePartial} rule described in \autoref{sec:theory:standardlib} is not optimized out because it is not surrounded by a $\gen{pick}$.


  \begin{table}[H]
    \begin{tabular}{|l|l|rr|}
      \hline
      {\bf Predicate} & {\bf Type} & \multicolumn{2}{|l|}{\bf Time (s)} \\\hline
      \code{$\mathbf{v}$ = 2} & \code{Nat} & $0.04$ & ($0.00$) \\ 
      \code{2 =\ $\mathbf{v}$} & \code{Nat} & $0.05$ & ($0.00$) \\ 
      \code{$\mathbf{v}$ = 2\ $\lor$\ $\mathbf{v}$ = 5} & \code{Nat} & $0.08$ & ($0.01$) \\ 
      \code{$\mathbf{v}$ = 2\ $\lor$\ $\mathbf{v}$ = 5\ $\land$ True} & \code{Nat} & $0.07$ & ($0.00$) \\ 
      \code{$\exists$ a, a = 3\ $\land$\ $\mathbf{v}$ = a + 1} & \code{Nat} & $0.05$ & ($0.00$) \\ 
      \code{5 <=\ $\mathbf{v}$\ $\land$\ $\mathbf{v}$\ <= 10} & \code{Nat} & $0.07$ & ($0.00$) \\ 
      \code{$\mathbf{v}$ > 5} & \code{Nat} & $0.06$ & ($0.00$) \\ 
      \code{$\mathbf{v}$ = 0\ $\lor$ lo <=\ $\mathbf{v}$\ $\land$\ $\mathbf{v}$ <= hi} & \code{Nat} & $0.12$ & ($0.01$) \\ 
      \code{isAllTwos\ $\mathbf{v}$ = true} & \code{List Nat} & $0.82$ & ($0.02$) \\ 
      \code{isAllTwosEvenLen\ $\mathbf{v}$ = true} & \code{List Nat} & $2.36$ & ($0.4$) \\ 
      \code{isEvenLen\ $\mathbf{v}$ = true} & \code{List Nat} & $2.15$ & ($0.04$) \\ 
      \code{isIncreasingByOne\ $\mathbf{v}$ = true} & \code{List Nat} & $1.46$ & ($0.03$) \\ 
      \code{List.length\ $\mathbf{v}$ = k} & \code{List Nat} & $1.86$ & ($0.03$) \\ 
      \code{isLengthKAllTwos k\ $\mathbf{v}$ = true} & \code{List Nat} & $3.38$ & ($0.05$) \\ 
      \code{isSortedBetween\ $\mathbf{v}$ (lo, hi) = true} & \code{List Nat} & $1.61$ & ($0.03$) \\ 
      \code{isAllEvens\ $\mathbf{v}$ = true} & \code{List Nat} & $0.92$ & ($0.02$) \\ 
      \code{isTrue\ $\mathbf{v}$ = true} & \code{List Nat} & $2.14$ & ($0.04$) \\ 
      \code{isUnique\ $\mathbf{v}$} & \code{List Nat} & \multicolumn{2}{l|}{FAIL in 1.9s} \\ 
      \code{hasDuplicates\ $\mathbf{v}$} & \code{List Nat} & \multicolumn{2}{l|}{FAIL in 1.75s} \\ 
      \code{isAllTwos\ $\mathbf{v}$ = true} & \code{Tree Nat} & $2.17$ & ($0.04$) \\ 
      \code{isBST\ $\mathbf{v}$ (lo, hi) = true} & \code{Tree Nat} & $1.53$ & ($0.04$) \\ 
      \code{isComplete\ $\mathbf{v}$ n = true} & \code{Tree Nat} & $1.57$ & ($0.03$) \\ 
      \code{isMaxDepth\ $\mathbf{v}$ n = true} & \code{Tree Nat} & $1.65$ & ($0.04$) \\ 
      \code{isIncreasingByOne\ $\mathbf{v}$ = true} & \code{Tree Nat} & $1.35$ & ($0.03$) \\ 
      \code{isNonempty\ $\mathbf{v}$ = true} & \code{Tree Nat} & $1.40$ & ($0.03$) \\ 
      \code{isGoodStack\ $\mathbf{v}$ n = true} & \code{Stack} & $2.76$ & ($0.06$) \\ 
      \code{isWellScoped\ $\mathbf{v}$ 0 = true} & \code{Term} & $1.64$ & ($0.03$) \\ 
      \code{isWellTyped\ $\Gamma$\ $\mathbf{v}$} & \code{Term} & $2.47$ & ($0.05$) \\ 
      \hline
      \code{lo <=\ $\mathbf{v}$\ $\land$\ $\mathbf{v}$\ <= hi} & \code{Nat} & $0.06$ & ($0.00$) \\ 
      \code{isAVL height lo hi\ $\mathbf{v}$ = true} & \code{Tree Nat} & $8.29$ & ($0.18$) \\ 
      \code{isRBTNoBST height\ $\mathbf{v}$ = true} & \code{Tree Nat} & $8.80$ & ($0.22$) \\ 
      \code{isRBT height lo hi\ $\mathbf{v}$ = true} & \code{Tree Nat} & $50.4$ & ($0.89$) \\ 
      \hline
    \end{tabular}
    \label{tab:app-eval}
  \end{table}

  \newpage

  \subsection{\cobbbench}

  \begin{table}[H]
    \begin{tabular}{|ll|r|r|}
      \hline
                      &            & {\bf Maximal Sketch } & {\bf Minimal Sketch} \\
      {\bf Predicate} & {\bf Type} & {\bf Time (s)}        & {\bf Time (s)} \\\hline
      \code{isSortedBetween\ $\mathbf{v}$ (lo, hi) = true} & \code{List Nat} & 0.58 & 5.58 \\
      \code{isAllEvens\ $\mathbf{v}$ = true} & \code{List Nat} & 2.56 & 10.35 \\
      \code{isTrue\ $\mathbf{v}$ = true} & \code{List Nat} & \multicolumn{2}{c|}{Insufficient Lemmas} \\
      \code{isIncreasingByOne\ $\mathbf{v}$ = true} & \code{List Nat} & 0.60 & 5.33 \\
      \code{isAllTwos\ $\mathbf{v}$ = true} & \code{List Nat} & 1.11 & 3.32 \\
      \code{isAllTwosEvenLen\ $\mathbf{v}$ = true} & \code{List Nat} & 0.69 & Timeout \\
      \code{isEvenLen\ $\mathbf{v}$ = true} & \code{List Nat} & 0.72 & Timeout \\
      \code{List.length\ $\mathbf{v}$ = k} & \code{List Nat} & 0.61 & 2.16 \\
      \code{isLengthKAllTwos k\ $\mathbf{v}$ = true} & \code{List Nat} & 0.67 & 2.70 \\
      \code{isUnique\ $\mathbf{v}$} & \code{List Nat} & 0.52 & 1.51 \\
      \code{hasDuplicates\ $\mathbf{v}$} & \code{List Nat} & 0.66 & 3.92\\
      \code{isMaxDepth\ $\mathbf{v}$ n = true} & \code{Tree Nat} & 1.17 & 8.29 \\
      \code{isComplete\ $\mathbf{v}$ n = true} & \code{Tree Nat} & 0.7 & 4.20 \\
      \code{isIncreasingByOne\ $\mathbf{v}$ = true} & \code{Tree Nat} & 1.54 & 191.74 \\
      \code{isNonempty\ $\mathbf{v}$ = true} & \code{Tree Nat} & 3.48 & 11.82 \\
      \code{isAllTwos\ $\mathbf{v}$ = true} & \code{Tree Nat} & 2.99 & 16.48 \\
      \code{isRBTNoBST height\ $\mathbf{v}$ = true} & \code{Tree Nat} & 39.62 & 64.9 \\
      \code{isRBT height lo hi\ $\mathbf{v}$ = true} & \code{Tree Nat} & \multicolumn{2}{c|}{Insufficient Lemmas} \\
      \code{isBST\ $\mathbf{v}$ (lo, hi) = true} & \code{Tree Nat} & 46.08 & 736.76 \\
      \code{isAVL height lo hi\ $\mathbf{v}$ = true} & \code{Tree Nat} & \multicolumn{2}{c|}{Insufficient Lemmas} \\
       \code{isWellTyped\ $\Gamma$\ $\mathbf{v}$} & \code{Term} & 5.6 & 577.86 \\
      \code{isWellScoped\ $\mathbf{v}$ 0 = true} & \code{Term} & \multicolumn{2}{c|}{Insufficient Lemmas} \\
      \code{isGoodStack\ $\mathbf{v}$ n = true} & \code{Stack} & \multicolumn{2}{c|}{Insufficient Lemmas} \\
      \hline
      \code{$\mathbf{v}$ = 2} & \code{Nat} & \multicolumn{2}{c|}{Immediate failure (no sketch)} \\ 
      \code{2 =\ $\mathbf{v}$} & \code{Nat} & \multicolumn{2}{c|}{Immediate failure (no sketch)} \\ 
      \code{$\mathbf{v}$ = 2\ $\lor$\ $\mathbf{v}$ = 5} & \code{Nat} & \multicolumn{2}{c|}{Immediate failure (no sketch)} \\ 
      \code{$\mathbf{v}$ = 2\ $\lor$\ $\mathbf{v}$ = 5\ $\land$ True} & \code{Nat} & \multicolumn{2}{c|}{Immediate failure (no sketch)} \\ 
      \code{$\exists$ a, a = 3\ $\land$\ $\mathbf{v}$ = a + 1} & \code{Nat} & \multicolumn{2}{c|}{Immediate failure (no sketch)} \\ 
      \code{5 <=\ $\mathbf{v}$\ $\land$\ $\mathbf{v}$ <= 10} & \code{Nat} & \multicolumn{2}{c|}{Immediate failure (no sketch)} \\ 
      \code{$\mathbf{v}$ > 5} & \code{Nat} & \multicolumn{2}{c|}{Immediate failure (no sketch)} \\ 
      \code{$\mathbf{v}$ = 0 $\lor$\ lo <=\ $\mathbf{v}$\ $\land$\ $\mathbf{v}$ <= hi} & \code{Nat} & \multicolumn{2}{c|}{Immediate failure (no sketch)}\\
      \code{lo <=\ $\mathbf{v}$\ $\land$\ $\mathbf{v}$ <= hi} & \code{Nat} & \multicolumn{2}{c|}{Immediate failure (no sketch)} \\
      \hline
    \end{tabular}
  \end{table}

  \newpage

\subsection{QuickChick}
\label{app:QC}

\newcommand{\cnatural}{\textcolor{dkgreen}{Natural}}
\newcommand{\coutofscope}{\textcolor{red}{Out-of-Scope}}
\newcommand{\cmerging}{\textcolor{blue}{Merging}}
\newcommand{\ctweak}{\textcolor{orange}{Tweaking}}

Classification of different benchmarks on the translation process of the \mainbench benchmarks.
\cnatural{} denotes that the natural direct translation worked, \cmerging{} denotes that they needed
to manually invoke merging to obtain an efficient generator, \ctweak{} denotes that the order of recursive premises matters for performance so tweaking might be required, \coutofscope{} is an existential
which is out of QuickChick's automation scope.

\begin{table}[H]
    \begin{tabular}{|l|c|}
      \hline
      {\bf Predicate} & {\bf Classification} \\ \hline
      \code{$\mathbf{v}$ = 2} & \cnatural \\
      \code{2 = $\mathbf{v}$} & \cnatural \\
      \code{$\mathbf{v}$ = 2\ $\lor$\ $\mathbf{v}$ = 5} & \cnatural \\
      \code{$\mathbf{v}$ = 2\ $\lor$\ $\mathbf{v}$ = 5\ $\land$ True} & \cmerging \\
      \code{$\exists$ a, a = 3\ $\land$\ $\mathbf{v}$ = a + 1} & \coutofscope \\
      \code{5 <=\ $\mathbf{v}$\ $\land$\ $\mathbf{v}$ <= 10} & \cmerging \\
      \code{$\mathbf{v}$ > 5} & \cnatural \\
      \code{$\mathbf{v}$ = 0\ $\lor$ lo <=\ $\mathbf{v}$\ $\land$\ $\mathbf{v}$ <= hi} & \cmerging \\
      \code{isAllTwos\ $\mathbf{v}$ = true} & \cnatural \\
      \code{isAllTwosEvenLen\ $\mathbf{v}$ = true} & \cmerging \\
      \code{isEvenLen\ $\mathbf{v}$ = true} & \cnatural \\
      \code{isIncreasingByOne\ $\mathbf{v}$ = true} & \cnatural \\
      \code{List.length\ $\mathbf{v}$ = k} & \cnatural \\
      \code{isLengthKAllTwos k\ $\mathbf{v}$ = true} & \cmerging \\
      \code{isSortedBetween\ $\mathbf{v}$ (lo, hi) = true} & \cmerging \\
      \code{isAllEvens\ $\mathbf{v}$ = true} & \cnatural \\
      \code{isTrue\ $\mathbf{v}$ = true} & \cnatural \\
      \code{isUnique\ $\mathbf{v}$} & \cnatural \\
      \code{hasDuplicates\ $\mathbf{v}$} & \cnatural \\
      \code{isAllTwos\ $\mathbf{v}$ = true} & \cnatural \\
      \code{isBST\ $\mathbf{v}$ (lo, hi) = true} & \cmerging \\
      \code{isComplete\ $\mathbf{v}$ n = true} & \cnatural \\
      \code{isMaxDepth\ $\mathbf{v}$ n = true} & \cnatural \\
      \code{isIncreasingByOne\ $\mathbf{v}$ = true} & \cnatural \\
      \code{isNonempty\ $\mathbf{v}$ = true} & \cnatural \\
      \code{isGoodStack\ $\mathbf{v}$ n = true} & \cnatural \\
      \code{isWellScoped\ $\mathbf{v}$ 0 = true} & \cnatural \\
      \code{isWellTyped\ $\Gamma$\ $\mathbf{v}$} & \ctweak \\
      \hline
      \code{lo <=\ $\mathbf{v}$\ $\land$\ $\mathbf{v}$ <= hi} & \cmerging \\
      \code{isAVL height lo hi\ $\mathbf{v}$ = true} & \cmerging \\
      \code{isRBTNoBST height\ $\mathbf{v}$ = true} & \cnatural \\
      \code{isRBT height lo hi\ $\mathbf{v}$ = true} & \cmerging \\
      \hline
    \end{tabular}
  \end{table}

  \newpage

  \section{Manually Written STLC Generator}\label{appendix:manual-stlc}

\begin{lstlisting}
  def genWellTyped (@$\Gamma$@ : List Ty) : Gen Term := by
    let @$\tau$@ <- arbTy
    Term.unfold
      (fun (@$\tau$@, @$\Gamma$@) => do
        pick
          (caseTy @$\tau$@
            (fun () =>
              -- $\tau$ = .unit
              pure TermF.unitStep)
            (fun @$\tau$@1 @$\tau$@2 () =>
              -- $\tau$ = .arrow $\tau$1 $\tau$2
              pure (TermF.absStep @$\tau$@1 (@$\tau$@2, @$\tau$@1 :: @$\Gamma$@))))
          (if (@$\Gamma$@.indexesOf @$\tau$@).length > 0 then
            pick
              (do
                let n <- elements (@$\Gamma$@.indexesOf .unit) (...)
                pure (TermF.varStep n))
              (do
                let @$\tau$@' <- arbTy
                pure (TermF.appStep (.arrow @$\tau$@' @$\tau$@, @$\Gamma$@) (@$\tau$@', @$\Gamma$@)))
          else do
            let @$\tau$@' <- arbTy
            pure (TermF.appStep (.arrow @$\tau$@' @$\tau$@, @$\Gamma$@) (@$\tau$@', @$\Gamma$@))))
      (@$\tau$@, @$\Gamma$@)
\end{lstlisting}

  \newpage

  \section{Synthesized Generator of Well-Typed Terms}\label{appendix:synthesized-genWellTyped}

\begin{lstlisting}
  def genWellTyped (@$\Gamma$@ : List Ty) : Gen Term := by
    let @$\tau$@ <- arbTy; Term.unfold
      (fun (@$\tau$@, @$\Gamma$@) => do
        let step <- caseTy @$\tau$@
          (fun () =>                                                   -- $\tau$ = .unit
            pick
              (pure TermF.unitStep)
              (if (@$\Gamma$@.indexesOf .unit).length > 0 then
                pick (do let n <- elements (@$\Gamma$@.indexesOf .unit) (...)
                      pure (TermF.varStep n))
                      (do let @$\tau$@' <- arbTy
                          pure (TermF.appStep (.arrow @$\tau$@' .unit) @$\tau$@'))
              else do
                let @$\tau$@' <- arbTy; pure (TermF.appStep (.arrow @$\tau$@' .unit) @$\tau$@')))
          (fun @$\tau$@1 @$\tau$@2 ()   =>                                           -- $\tau$ = .arrow $\tau$1 $\tau$2
            if (@$\Gamma$@.indexesOf (.arrow @$\tau$@1 @$\tau$@2)).length > 0 then
              pick (do let n <- elements (@$\Gamma$@.indexesOf (.arrow @$\tau$@1 @$\tau$@2)) (...)
                    pure (TermF.varStep n))
                    (pick
                      (pure (TermF.absStep @$\tau$@1 @$\tau$@2))
                      (do let @$\tau$@' <- arbTy
                          pure (TermF.appStep (.arrow @$\tau$@' (.arrow @$\tau$@1 @$\tau$@2)) @$\tau$@')))
            else
              pick (pure (TermF.absStep @$\tau$@1 @$\tau$@2))
                    (do let @$\tau$@' <- arbTy
                        pure (TermF.appStep (.arrow @$\tau$@' (.arrow @$\tau$@1 @$\tau$@2)) @$\tau$@')))
        match step with
        | TermF.unitStep => pure TermF.unitStep
        | TermF.varStep n => pure (TermF.varStep n)
        | TermF.absStep @$\tau$@ b => pure (TermF.absStep @$\tau$@ (b, @$\tau$@ :: @$\Gamma$@))
        | TermF.appStep b1 b2 => pure (TermF.appStep (b1, @$\Gamma$@) (b2, @$\Gamma$@))) (@$\tau$@, @$\Gamma$@)
\end{lstlisting}

  \newpage

  \section{Synthesized Generator of AVL Trees}\label{appendix:synthesized-genAVL}

\begin{lstlisting}
def genAVL (height lo hi : Nat) : Gen (Tree Nat) :=
  Tree.unfold
    (fun x => do
      let __do_lift <-
        if x.snd.snd = 0 then pure TreeF.leaf
          else
            if Nat.pred x.snd.snd = 0 then
              if h : decide (x.snd.fst.fst <= x.snd.fst.snd) = true then
                pick (pure TreeF.leaf) do
                  let a <- choose x.2.1.1 x.2.1.2 (s_between_partial._proof_1 h)
                  pure (TreeF.node (PUnit.unit, PUnit.unit) a (PUnit.unit, PUnit.unit))
              else pure TreeF.leaf
            else
              assume (decide (x.snd.fst.fst <= x.snd.fst.snd)) fun h => do
                let a <- choose x.2.1.1 x.2.1.2 (s_between_partial._proof_1 h)
                pure (TreeF.node (PUnit.unit, PUnit.unit) a (PUnit.unit, PUnit.unit))
      match __do_lift with
        | TreeF.leaf => pure TreeF.leaf
        | TreeF.node bl x_1 br =>
          pure
            (TreeF.node (bl, (x.2.1.1, x_1 - 1), x.2.2 - 1) x_1
              (br, (x_1 + 1, x.2.1.2), x.2.2 - 1)))
    ((PUnit.unit, PUnit.unit), (lo, hi), height)
\end{lstlisting}

  \newpage

  \section{Manually Written Generators of AVL Trees}\label{appendix:manual-genAVL}
\begin{lstlisting}
/-
Differences:
- Remove two extra units in collector.
- Nicer match on height to reduce some duplication.
- Generator is technically total now; this requires insight about the total
  number of values that can appear in a tree of height k.
-/
def genAVL_manual' (height lo hi : Nat) : Gen (Tree Nat) :=
  -- Guarantee that there are enough values in the range, given the height.
  assume (hi - lo > 2 ^ height) fun _ =>
    Tree.unfold
      (fun (lo, hi, height) => do
        match height with
        | 0 => pure TreeF.leaf
        | 1 =>
            pick (pure TreeF.leaf)
              (assume (lo <= hi) fun h => do  -- Will always succeed.
                -- Choose values so we never truncate the range to be too small.
                let a <-
                  choose
                    (lo + 2 ^ (height - 1))
                    (hi - 2 ^ (height - 1)) (by ...)
                pure (TreeF.node (lo, a - 1, height - 1) a (a + 1, hi, height - 1)))
        | height' + 1 => do
          assume (lo <= hi) fun h => do -- Will always succeed.
            -- Choose values so we never truncate the range to be too small.
            let a <-
              choose
                (lo + 2 ^ (height - 1))
                (hi - 2 ^ (height - 1)) (by ...)
            pure (TreeF.node (lo, a - 1, height - 1) a (a + 1, hi, height - 1)))
      (lo, hi, height)

/-
Differences: Entirely different approach; relies on AVL.insert being correct.
-/
def genAVL_manual'' : Gen (Tree Nat) :=
  -- Generate list of arbitrary Nats
  let values <-
    List.unfold (fun () =>
      pick
        (pure (ListF.nil))
        (do let a <- arbNat; pure (ListF.cons a ())))
    ()
  -- Insert all values into an empty AVL tree.
  pure (List.fold (fun x t => AVL.insert t x) AVL.empty)
\end{lstlisting}

  \section{More Side-by-Side Generator Comparisons}\label{appendix:side-by-side}

  \subsection{One Or In Range}
\begin{lstlisting}
def genOneOrInRange (lo hi : Nat) : Gen Nat :=
  if h : decide (lo <= hi) = true then
    pick (pure 0) (choose lo hi (s_between_partial._proof_1 h))
  else
    pure 0

/-
Differences:
- Simplify proof for choose.
-/
def genOneOrInRange_manual (lo hi : Nat) : Gen Nat :=
  if h : lo <= hi then
    pick (pure 0) (choose lo hi (by omega))
  else
    pure 0
\end{lstlisting}

  \subsection{Complete Tree}
\begin{lstlisting}
def genCompleteTree (n : Nat) : Gen (Tree Nat) :=
  Tree.unfold
    (fun x =>
      if x.snd = 0 then pure TreeF.leaf
      else do
        let a <- arbNat
        pure (TreeF.node ((), x.2 - 1) a ((), x.2 - 1)))
    ((), n)

/-
Differences:
- Remove extra unit in collector.
-/
def genComplete_manual (n : Nat) : Gen (Tree Nat) :=
  Tree.unfold
    (fun height =>
      if height = 0 then
        pure TreeF.leaf
      else do
        let a <- arbNat
        pure (TreeF.node (height - 1) a (height - 1)))
    n
\end{lstlisting}

  \newpage

  \subsection{Sorted Between}
\begin{lstlisting}
  def genSortedBetween (lo hi : Nat) : Gen (List Nat) :=
  List.unfold
    (fun x =>
      if h : decide (x.snd.fst <= x.snd.snd) = true then
        pick (pure ListF.nil) do
          let a <- choose x.2.1 x.2.2 (s_between_partial._proof_1 h)
          pure (ListF.cons a (PUnit.unit, a, x.2.2))
      else
        pure ListF.nil)
    (PUnit.unit, lo, hi)

/-
Differences:
- Simplify proof for choose.
- Remove extra unit in collector.
-/
def genSortedBetween_manual (lo hi : Nat) : Gen (List Nat) :=
  List.unfold
    (fun (lo, hi) =>
      if h : lo <= hi then
        pick
          (pure ListF.nil)
          (do
            let a <- choose lo hi (by omega)
            pure (ListF.cons a (a, hi)))
      else
        pure ListF.nil)
    (lo, hi)
\end{lstlisting}

  \newpage

  \subsection{Length K, All Twos}
\begin{lstlisting}
def genLengthKAllTwos (k : Nat): Gen (List Nat) :=
  List.unfold
    (fun x =>
      if x.fst.fst = 0 then pure ListF.nil
      else pure (ListF.cons 2 ((Nat.pred x.1.1, PUnit.unit), PUnit.unit, PUnit.unit)))
    ((k, PUnit.unit), PUnit.unit, PUnit.unit)

/-
Differences:
- Remove two extra units in collector.
-/
def genLengthKAllTwos_manual (k : Nat): Gen (List Nat) :=
  List.unfold
    (fun len =>
      if len = 0 then
        pure ListF.nil
      else
        pure (ListF.cons 2 (len - 1)))
    k
\end{lstlisting}

  \end{document}